\documentclass[twocolumn,pre,showpacs,superscriptaddress]{revtex4}
\usepackage{graphicx}%
\usepackage{psfrag}
\usepackage{dcolumn}
\usepackage{amsmath}
\usepackage{amssymb}
\usepackage{latexsym}
\usepackage{hyperref} 
\begin{document}

\def\bea{\begin{eqnarray}}
\def\eea{\end{eqnarray}}
\def\beq{\begin{equation}}
\def\eeq{\end{equation}}
\def\f{\frac}
\def\k{\kappa}
\def\sx{\sigma_{xx}}
\def\sy{\sigma_{yy}}
\def\sxy{\sigma_{xy}}
\def\e{\epsilon}
\def\ve{\varepsilon}
\def\ex{\epsilon_{xx}}
\def\ey{\epsilon_{yy}}
\def\exy{\epsilon_{xy}}
\def\be{\beta}
\def\D{\Delta}
\def\h{\theta}
\def\t{\tau}
\def\r{\rho}
\def\a{\alpha}
\def\s{\sigma}
\def\kb{k_B}
\def\la{\langle}
\def\ra{\rangle}
\def\nn{\nonumber}
\def\bu{{\bf u}}
\def\bn{\bar{n}}
\def\br{{\bf r}}
\def\up{\uparrow}
\def\dn{\downarrow}
\def\S{\Sigma}
\def\dg{\dagger}
\def\d{\delta}
\def\p{\partial}
\def\l{\lambda}
\def\G{\Gamma}
\def\o{\omega}
\def\g{\gamma}
\def\kv{\bar{k}}
\def\ha{\hat{A}}
\def\hv{\hat{V}}
\def\hg{\hat{g}}
\def\hG{\hat{G}}
\def\hTT{\hat{T}}
\def\noi{\noindent}
\def\a{\alpha}
\def\d{\delta}
\def\p{\partial} 
\def\r{\rho}
\def\xv{\vec{x}}
\def\rv{\vec{r}}
\def\fv{\vec{f}}
\def\ov{\vec{0}}
\def\vv{\vec{v}}
\def\la{\langle}
\def\ra{\rangle}
\def\e{\epsilon}
\def\o{\omega}
\def\n{\eta}
\def\g{\gamma}
\def\th{\hat{t}}
\def\uh{\hat{u}}
\def\break#1{\pagebreak \vspace*{#1}}
\def\f{\frac}
\def\hf{\frac{1}{2}}
\def\uu{\vec{u}}
\bibliographystyle{prsty}
\title{A model of fasciculation and sorting in mixed populations of axons}
\author{Debasish Chaudhuri}
\altaffiliation{
Current address:
FOM Institute AMOLF, Science Park 104, 1098 XG, Amsterdam, The Netherlands}
\email{d.chaudhuri@amolf.nl}
\affiliation{
Max Planck Institute for the Physics of Complex Systems, 
N{\"o}thnitzer Strasse 38, 
01187 Dresden, Germany
}

\author{Peter Borowski}
\email{peterphysik@gmail.com}
\affiliation{
Department of Physics, Indian Institute of Technology Madras, Chennai 600036, India
}

\author{Martin Zapotocky}
\email{zapotocky@biomed.cas.cz}
\affiliation{
Institute of Physiology,
Academy of Sciences of the Czech Republic,
Videnska 1083,
14220 Praha 4,
Czech Republic
}

\date{\today}

\begin{abstract}
We extend a recently proposed model 
(Chaudhuri {\em et al.}, EPL {\bf 87}, 20003 (2009)) 
aiming to describe the formation of fascicles of axons during
neural development. The growing axons are represented as 
paths of interacting directed random walkers in two spatial dimensions.
To mimic turnover of axons, whole paths are removed and new walkers are 
injected with specified rates.
In the simplest version of the model, we use strongly adhesive short-range
inter-axon interactions that are identical for all pairs of axons. 
We generalize the model to adhesive interactions of finite strengths
and to multiple types of axons with type-specific interactions.
The dynamic steady state is characterized by the  position-dependent
distribution of fascicle size and fascicle composition.  With distance
in the direction of axon growth, the mean fascicle size and emergent
time scales grow  monotonically, while the degree of sorting of 
fascicles by axon type  has a maximum  at a finite distance. 
To understand the emergence of slow time scales, we develop an analytical
framework to analyze the interaction between neighboring fascicles.

\end{abstract}

\maketitle

\section{Introduction}
Reaction-diffusion phenomena arise in diverse fields such  as 
physical chemistry~\cite{phys-chem}
or developmental biology~\cite{math-bio}.   
In certain reaction-diffusion systems, the process of 
path aggregation occurs, in which preferred paths of the 
diffusing elements are established and evolve in time.
The path aggregation process is found in  diverse 
realm of nature, e.g., in
formation of insect  pheromone trails~\cite{ant1, ant2, pheromone} and
human walking trails~\cite{human,helbing}, in
aggregation of trails of liquid droplets moving down a window pane, in
river basin formation~\cite{scheidegger,river-book}, etc.

One class of the mathematical models in which path aggregation processes have been studied 
is the active-walker models~\cite{Kayser1992,helbing} 
in which each walker while moving through the system
changes the surrounding environment locally, which in turn influences the later 
walkers. The ant trail formation is an example of  such a process~\cite{ant1,ant2}.
An ant leaves a chemical trail of pheromones on its path which the other
ants can sense and follow.  Evaporation of pheromone leads to an aging of these trails.
Similarly, the mechanism of  human and animal trail formation is mediated by the
deformation of vegetation that generates an interaction between 
earlier and later walkers~\cite{human,helbing}. 
This deformation, and therefore its impact, decays continuously with time~\cite{helbing}.

In a recent Letter~\cite{our_epl}  we analyzed the dynamics of path aggregation using
a simple model that belongs to the class of active walker systems 
discussed above. 
In contrast to the active-walker models, in our model the individual paths 
do not age gradually, but rather maintain their full identity until they are abruptly removed from 
the system. This particular rule for path aging was chosen to allow
application of our model to the process of axon fasciculation 
(formation of axon bundles~\cite{whyfascicle}),  
which we discuss below and more in detail in Sec.~\ref{biomot}. 

In order to develop neuronal connections, 
sensory neurons born in peripheral tissues project their axons 
(long tubular part of the neuron cell that conducts electrical excitations) 
towards target regions in the brain. 
Frequently, multiple axons come together to form axon fascicles, 
and may sort according to the cell type of the neuron to which the 
axon belongs.
This fascicle formation and sorting can be driven by
inter-axon interactions leading to, e.g.,  a pre-target spatial map
in the mammalian olfactory system~\cite{Bozza2009,imai09,Miller2010}.

%

In our model, the axons are represented as directed random walks in two spatial dimensions.
In Ref.~\cite{our_epl}  we formulated and analyzed the simplest version of the model, in which all axons belong to the same type and have strong adhesive interactions, so that each newly growing axon encountering an existing fascicle will join the fascicle and never detach. In the presence of axon turnover (aging of the paths), a steady state characterized by a distribution of fascicle sizes~\cite{our_epl} is eventually established. The focus of Ref.~\cite{our_epl} was on the analysis of the surprisingly long time scales that emerge from this simple dynamics.

In the current paper, we significantly extend this theoretical analysis. We develop an analytical description of the dynamics of two neighboring fascicles, and show how the slowest mode of their interaction gives raise to the slow time scales observed in Ref. ~\cite{our_epl}. We also systematically discuss the limited analogies that can be made between our 2-dimensional model and 1-dimensional models of particle coalescence~\cite{Avraham1990}, aggregation~\cite{redner}, and chipping~\cite{mustansir,Rajesh2001}. These analogies are useful for the understanding of stationary quantities of our model such as the distribution of fasicle sizes and the distribution of inter-fascicle separations.

The main contribution of the current paper, however, is to generalize the previous model of Ref.~\cite{our_epl}  to attractive interactions of finite strength (so that detachment of axons from fascicles is possible) and to multiple axon types with type-specific interactions. Such a generalization is necessary to allow the biological application of the model.

In the following section we give a detailed 
biological motivation for the model we consider.
In Sec.~\ref{model} we introduce the model and the Monte Carlo (MC)
simulation scheme that we use to investigate its properties numerically. 
Followed by this, in Sec.~\ref{overview}, 
we give a brief  overview of guiding concepts that will recur in
the rest of the paper. 
In Sec.~\ref{onesp}, we present a detailed analysis of the system 
containing axons of a single type that follows the ``always attach, never detach" rule. 
We extend the numerical results of Ref.~\cite{our_epl} for the properties of the steady 
state and for the emerging time scales. We review the analytical framework of 
single-fascicle dynamics, developed in Ref.~\cite{our_epl}, and significantly extend it 
by deriving results for the interaction dynamics of two neighboring fascicles.
In Sec.~\ref{onespD} we numerically study the effects of 
non-vanishing detachment rates of axons from a fascicle.
{
In Sec.~\ref{1dcomp} we discuss some limited analogies of our model to one
dimensional aggregation and coalescence processes.}
In Sec.~\ref{2sp} we discuss the sorting of fascicles by axon types in a system containing 
two types of axons, the simplest manifestation of a mixed population of axons. Finally,
we provide a summary of main results in Sec.~\ref{sum} and   
conclude in Sec.~\ref{conc}  by discussing the outlook for biological applications of our model. 

\section{Biological motivation}
\label{biomot}
{
Sensory neurons located in peripheral tissues connect to more central 
locations of the nervous system via axons~\cite{Shepherd1994}. 
}
During  the development of an organism, axons of newly maturing 
sensory neurons must establish connections to the proper 
location.  
Axon growth is initiated at the soma (main cell body) of each neuron, 
and  proceeds with a typical rate of extension  
$100 \mathrm{\mu m/h}$~\cite{Honig98}. 
The direction of growth is controlled by the dynamic {\it growth cone} 
structure at the tip of the axon.  
The growth cone probes the environment in its vicinity, and can 
detect gradients of spatially distributed chemical signals. 
In the absence of strong directional signals, the path 
of the growth cone is highly stochastic~\cite{katz1985,Maskery2005}, while in the 
presence of appropriate guidance cues, the direction of motion 
becomes strongly biased. 
{
The overall direction of axon growth may be guided by spatial gradients of
}
chemical cues generated by the target.
A number of distinct molecular guidance cues 
that influence neuronal development have been identified in recent 
years~\cite{mclaughlin05,Luo07}, 
and the response of the growth cone to graded 
cues has been studied theoretically~\cite{goodhill98,gierer1998,AvanOoyen2}. 
In this work, we do not directly model the axon guidance by graded
chemical cues, but subsume their influence into the setup of our model by
giving all axons a common preferred growth direction.

In this article, we study the 
collective effects that arise from {\em direct local interactions} 
among the growing axons. When such interactions are attractive, 
the growth cone of a newly growing axon tends to follow the 
tracks (i.e., the axon shafts) of older axons. The strength of 
this interaction is governed by the type and expression level of 
the relevant cell adhesion molecules~\cite{Honig98, Wolman07}. 
The resulting dynamics can lead to selective formation of fascicles 
of axons~\cite{Goodman1984,Lin94,Pittman2008}, a common and essential phenomenon in the 
developing nervous systems. 

An additional important aspect included in our model is 
that of {\em neuronal turnover}. During development, a significant 
portion of sensory neurons with fully grown axons may die, and be 
replaced by younger sensory neurons which attempt a new connection 
to the brain. 
For example, up to $80 \%$ of retinal ganglion cell axons are 
lost during the development of the visual system in the cat~\cite{Williams1986,cat}.  
In the mammalian olfactory system, both 
neuronal birth and death persist throughout the life of the animal, 
leading to a dynamical steady state pattern of connectivity. In 
particular, the average lifetime of an olfactory sensory neuron in the 
mouse is of the order of 1--2 months~\cite{Nakatani03}, which is 
less than one tenth of the mouse's lifespan.

To motivate the introduction of multiple types of axons into our model, we
now briefly discuss the intricate connectivity pattern of the mammalian
olfactory system, 
{
which implements the sense of smell.
}
In the mouse, the adult nasal epithelium  contains 
approximately $10^{6}$ olfactory sensory neurons, which send their 
axons through the olfactory tract to the olfactory bulb in the 
forebrain. 
Remarkably, the sensory neurons belong to approximately 
1200 distinct types~\cite{Mombaerts,Bozza2009}, and the axons of 
each type connect to a distinct 
neuronal structure, a glomerulus, on each olfactory bulb~\cite{Vassar1994, Mombaerts}.
Such precise connectivity is fully established only after several 
turnover periods, while in newborn mice, split glomeruli and 
glomeruli that mix several axon types are often 
observed~\cite{Nakatani03, Zou04}. 

{
In olfactory sensory neurons, elegant genetic analysis shows that the axonal type is 
determined by the expression of a specific odorant receptor gene~\cite{Mombaerts,Malnic,Serizawa06,homotypic2}.
}
Physiological experiments on mice show that the expression of
specific types of cell adhesion molecules, that dictates the strength of
adhesive forces between axons, is strongly correlated with
this axonal type~\cite{Serizawa06, Chehrehasa2006}. 
A wide range of strengths of interactions between axons 
may be generated through combinatorial expression of 
multiple types of cell adhesion molecules. 

\begin{figure*}[t]
\includegraphics[height=5.5cm]{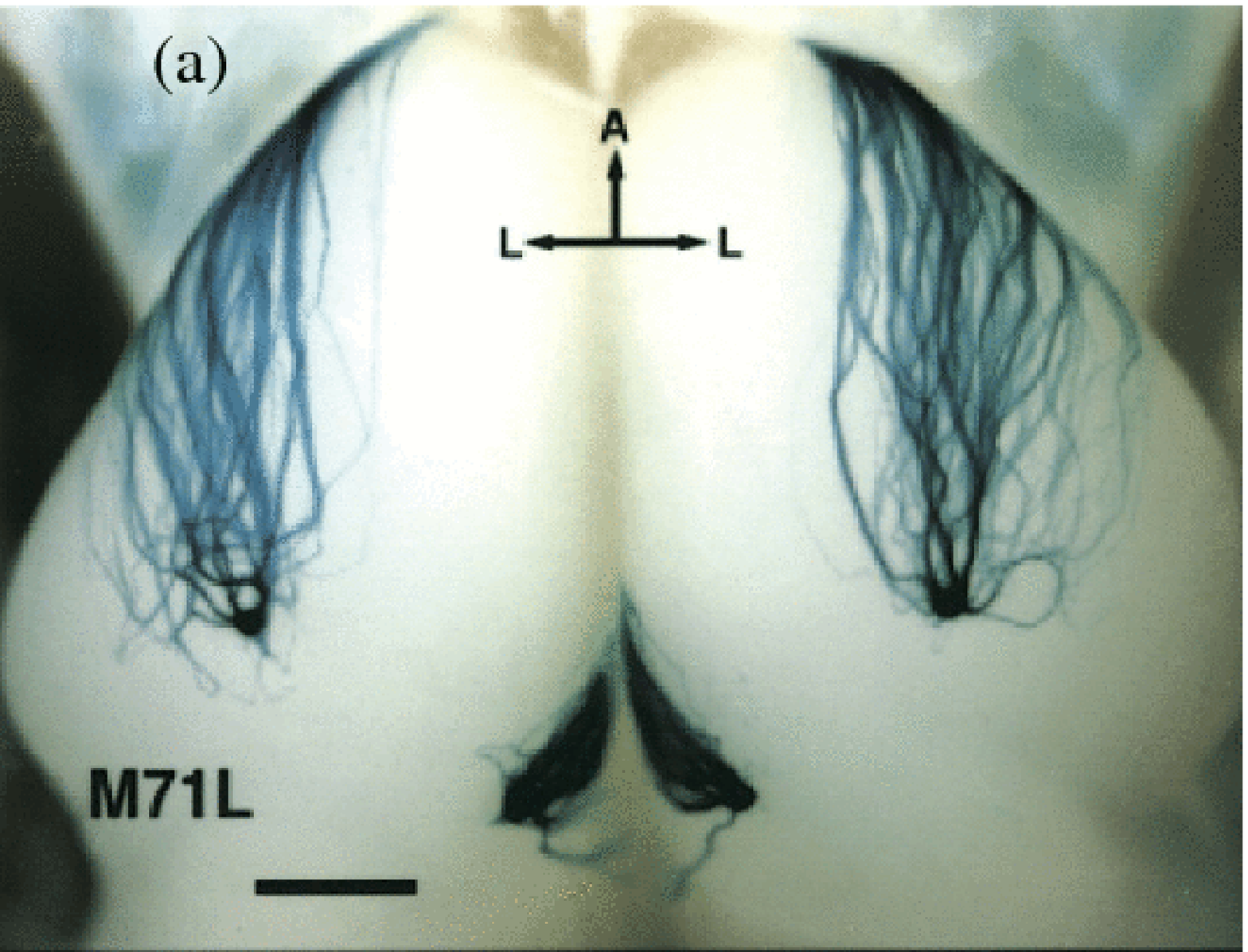}
\hskip 1cm
\includegraphics[height=5.5cm]{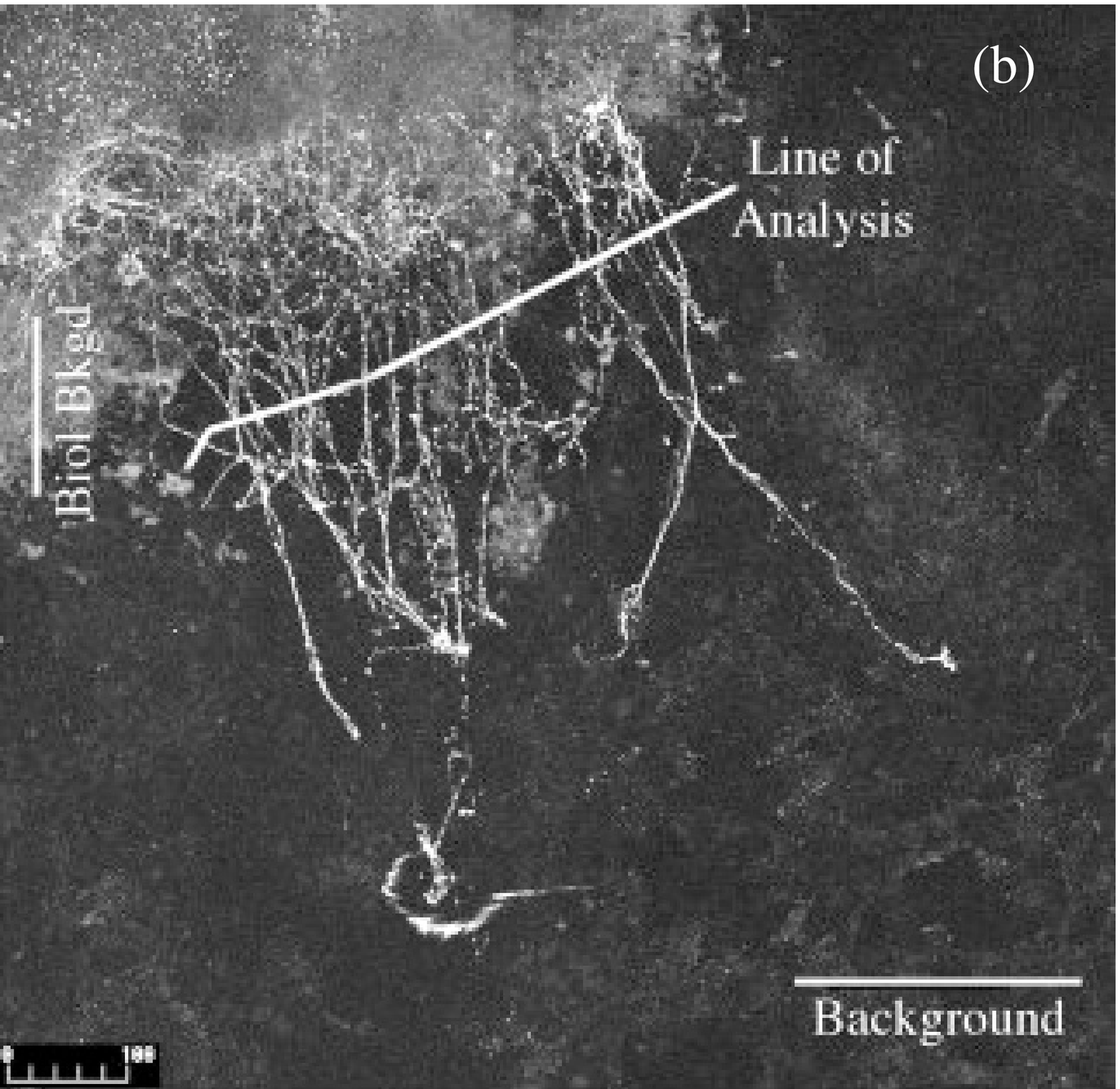}
\caption{(Color online) 
(a)~Axons of olfactory sensory neurons of a specific type (M71) growing in the surface layer of the mouse olfactory bulbs. Scale bar $= 500\, \mu$m. The axons emerge (top) from behind the olfactory bulbs, and grow towards the bottom. Note fasciculation as well events of detachments of axons from fascicles. Figure adapted from Ref.~\cite{homotypic2}. (b)~Axon growth (top to bottom) and fasciculation observed in explant culture of rat olfactory epithelium on a laminin-coated coverglass. Scale bar $= 100\, \mu$m. Axon type is not distinguished. Figure adapted from Ref.~\cite{hamlin}.
}
\label{bio_motiv}
\end{figure*}
{
In Fig.~\ref{bio_motiv} we show configurations of olfactory axons as observed in  
{\em in vivo}~\cite{homotypic2} and {\em in vitro}~\cite{hamlin} experiments. 
Fig.~\ref{bio_motiv}(a) shows axons growing in the surface layer of the left and right 
olfactory bulbs of a genetically modified mouse (Fig.1(L) of Ref.~\cite{homotypic2}). 
 Only axons belonging to one type of olfactory sensory neurons (expressing the M71 
 receptor gene) are labeled; axons of other types are present but not visible. 
 The axons progressively fasciculate and terminate in a glomerulus visible in the 
 center of each half-image. Fig.~\ref{bio_motiv}(b) shows  fasciculation of axons growing 
 from an explant of the rat olfactory epithelium  (Fig.7(a) of Ref.~\cite{hamlin}).    
 In this case, the fluorescent labeling does not distinguish the axonal type, and 
 (with a high probability) the visible axons belong to multiple types.
}

Our model aims to provide a quantitative framework for evaluating the
contribution of axon-axon interactions to 
{
the formation of patterns}
described above.
The presence of turnover and multiple axon types
in our model distinguishes our work from 
previous theoretical studies of axon fasciculation~\cite{AvanOoyen1,Hentschel1999}. 
Our implementation of the individual axon dynamics is particularly simple,
to allow us to concentrate on collective effects arising from interactions
within a population of axons.
\begin{figure}[h]
\begin{center}
\psfrag{(a)}{$(a)$}
\psfrag{(c)}{$(b)$}
\psfrag{x}{$x$}
\psfrag{y}{$y$}
\includegraphics[width=5.7cm]{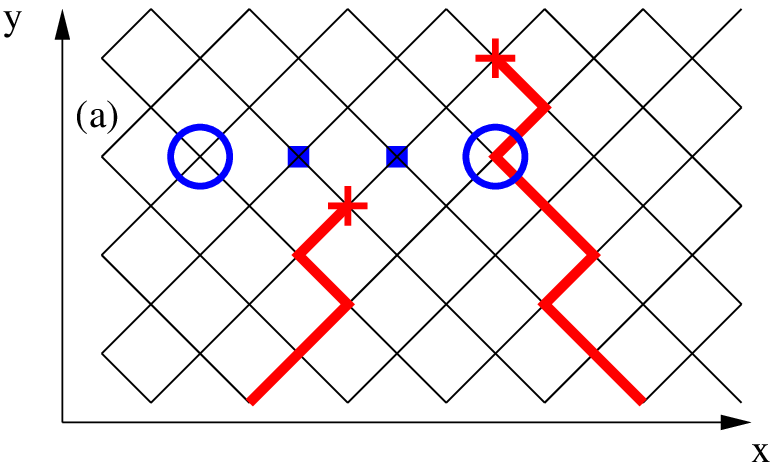}  
\includegraphics[width=6.4cm] {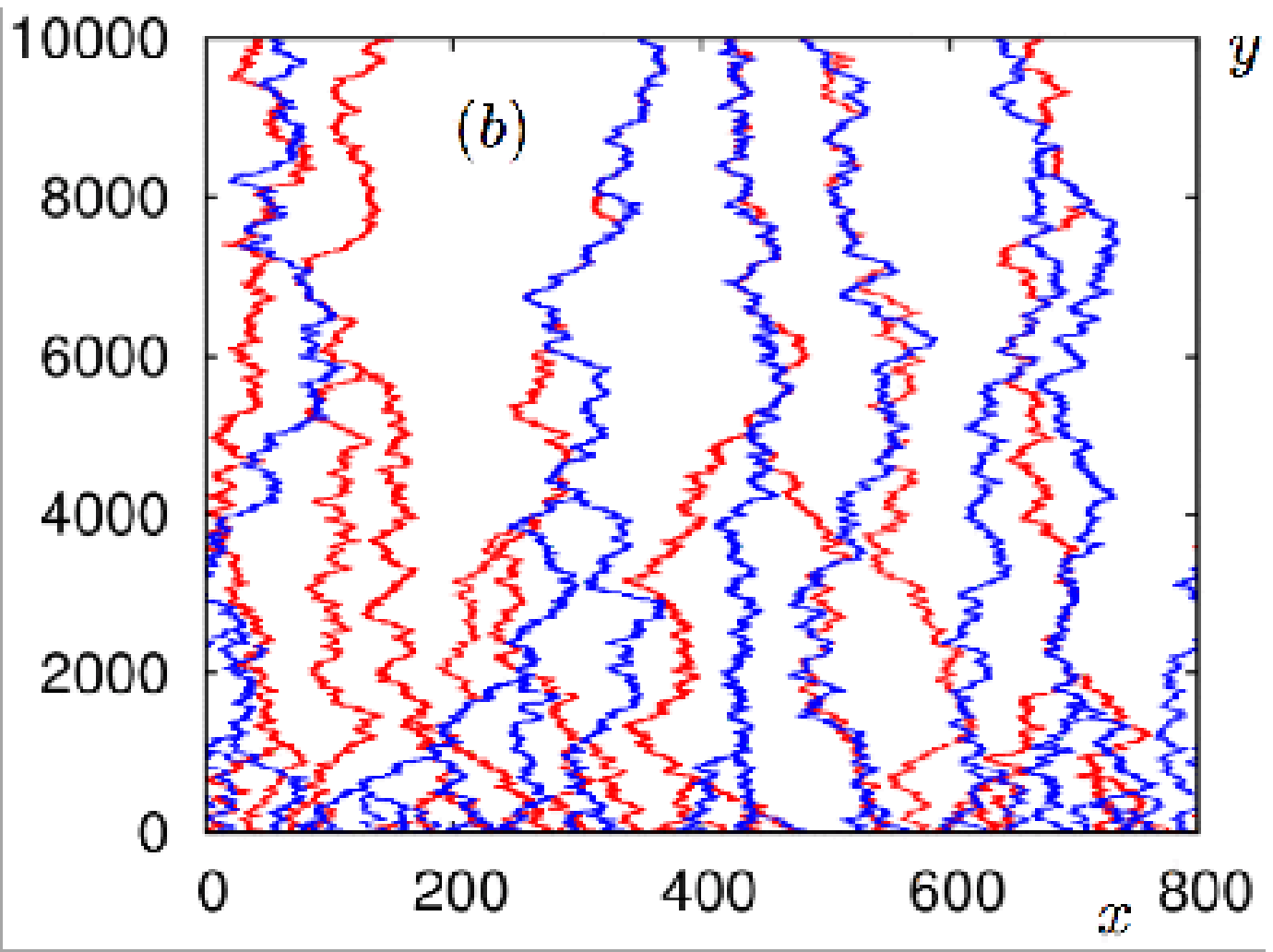}  
\psfrag{D}{$D$}
\psfrag{E}{$E$}
\psfrag{y}{$y$}
\psfrag{(b)}{$(c)$}
\includegraphics[width=8.3cm]{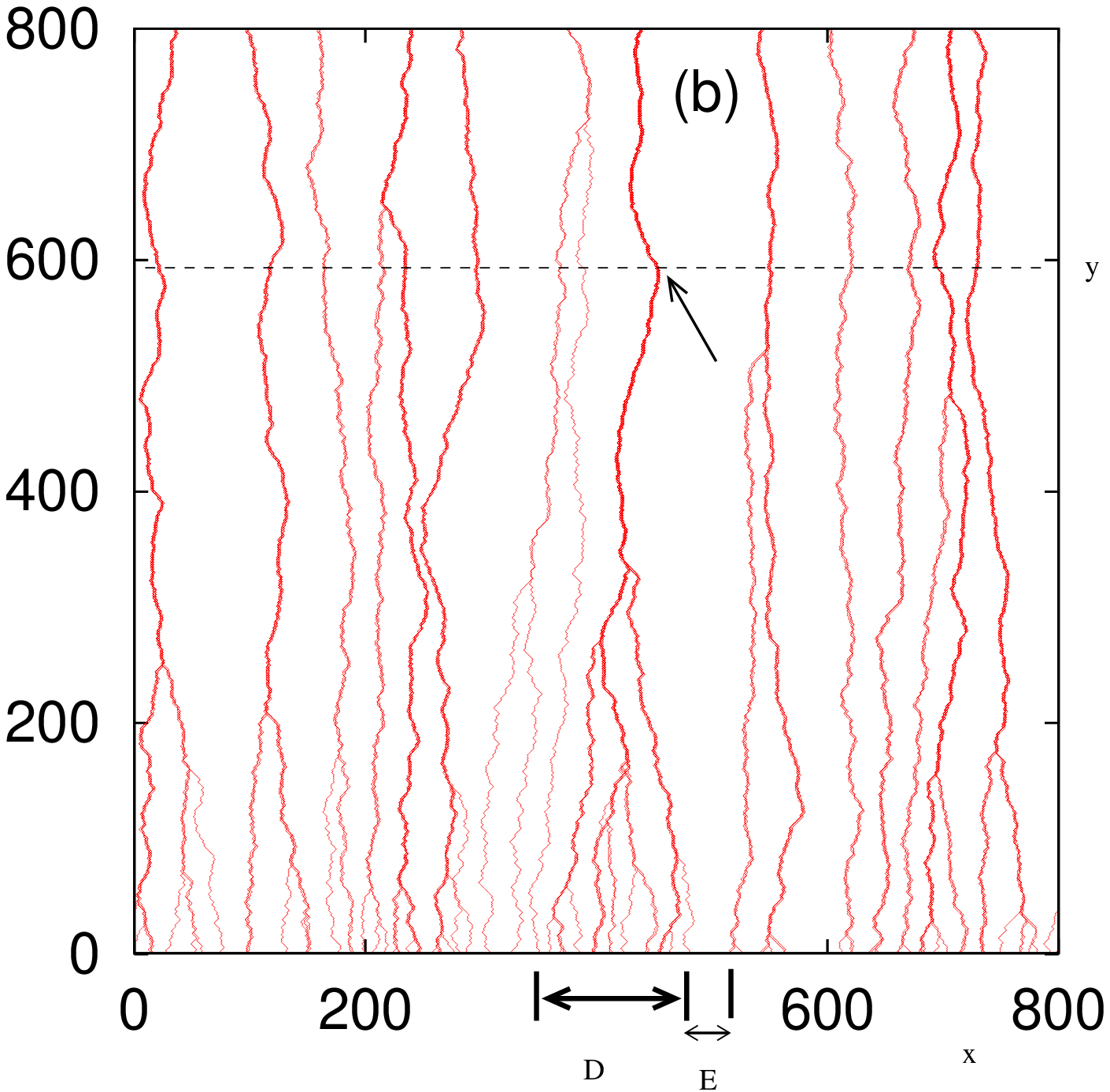} 
\end{center}
\caption{(Color online)
 ($a$)~Interacting directed random walks on a tilted square lattice. 
A random walker
($+$) represents a growth cone. For one walker, the possible future sites 
($\Box$) and their nearest neighbors ($\circ$) are marked. The trail of 
a walker (line) models an axon shaft.
($b$)~A typical late-time configuration ($t=25T$) of a system of axons belonging to two 
different types, $r$ (red) and  $b$ (blue).
The strength of the homotypic interaction is $E_h=-4$ and that of 
the heterotypic interaction is $E_o=-0.1$.
The mean numbers of $r$ and $b$ axons at $y=0$ are $N^r_0=N^b_0=50$ and
the system size is $L=800$. 
($c$)~A typical late-time configuration ($t=25T$) in a system 
with a single type of axons undergoing energy-minimizing dynamics with system parameters
$L=800$, $N_0=100$.  
For the fascicle identified at  $y=600$ (arrow), $D$ indicates its
basin 
and $E$ is the inter-basin free space.}
\label{conf}
\end{figure}
\section{Model and numerical implementation}
\label{model}

\subsection{Setup and interactions}
\label{setup}
In our model,
each growing axon is represented by a directed random walk in two spatial 
dimensions (Fig.~\ref{conf}$(a)$). The random walkers (representing the growth cones) are initiated at the epithelium 
($y=0$, random even $x$) with a birth rate 
$\a$, and move towards the bulb (large $y$) with constant velocity $v_y=1$. 
In the case of multi-type systems, a type is assigned to each newly initiated random walker (specifically in the simulations of Sec.~\ref{2sp}, the type is decided randomly with equal probability for each of the two types).
The trail generated by a random walker (growth cone) is regarded as
an axon shaft. A forward moving directed random walker (growth cone) interacts
with trails (axon shafts) of other walkers.
In the numerical implementation on a tilted square lattice, at each time step 
the growth cone at ($x,y$) can move to ($x-1,y+1$) (left) or ($x+1,y+1$) 
(right). The probability $p_{\{L,R\}}$ to move left/right is evaluated based 
on the axon occupancies at the ($x\pm 1,y+1$)  
sites and their nearest neighbors ($x\pm 3,y+1$) (see Fig.~\ref{conf}$(a)$). 
At a given $y$, two axons are considered to be part of the same fascicle 
if they are not separated by any unoccupied sites (i.e., they are not separated by more than two lattice spacings).

We assume a short-range attractive interaction between each growth cone and the close-by axon shafts. The range of interaction (two lattice spacings in our model) corresponds biologically to the range of extension of sensory filopodia from the growth cone (of the order of $10 \mu{\rm m}$). The attractive interaction is mediated by cell adhesion between  the growth cone and the axon shafts. We assume that the interactions are additive and type-specific.  For a given growth cone, the model assumes a weak nearest neighbor attraction
$E_o<0$ if the neighboring axon shaft is of a different type
and a stronger attraction $E_h\,( <E_o)$ if the neighboring axon shaft is of
the same type. 
In each time step, a growth cone at $(x,y)$ attempts a Monte Carlo move to the left $(x-1,y+1)$ and to the right  $(x+1,y+1)$ with probabilities $1/2$. The moves are accepted with probabilities
$$p_L=\mbox{min}[1,\exp(-\d E_l)]$$ 
(or $p_R=\mbox{min}[1,\exp(-\d E_r)]$) where 
\bea
\d E_l &=& [n_h(x-3,y+1)-n_h(x+3,y+1)] E_h \nn\\
&+& [n_o(x-3,y+1)-n_o(x+3,y+1)] E_o\nn
\eea
($\d E_r=-\d E_l$)
and the occupancy number $n_h$ denotes  the number of axons belonging to the same type as the growth cone, while $n_o$ is the number of axons of other types. 
Notice that in calculating the difference in energy, the occupancy of the positions
$(x\pm 1,y+1)$ does not appear, as their contributions to the energy cost mutually cancel. 
Periodic boundary conditions are used in the $x$-direction. 
In this kinetic MC scheme,
we used parallel updates of all random walkers in each MC step.

We now clarify the relation of the general model described above to the model studied by us in Ref.~\cite{our_epl}. 
In this simple version of the model, all axons belonged to the same type and
the interaction between them was governed by the ``always attach, never detach'' rule, which is the ``zero temperature" 
version of the above-mentioned general MC scheme, i.e., the dynamics was a pure energy minimization process:
$p_L = 1$ ($p_R=1$) when 
$\d E_l <0$ ($\d E_r <0$);
$p_L = p_R = 1/2$ in all other cases. This is in contrast to the ``finite temperature" dynamics of the general model, in which there is a non-zero rate for the detachment of growth cones from fascicles. 
Ref.~\cite{our_epl} used  sequential updates, in contrast to 
parallel updates used in this paper.
An additional difference is that in Ref.~\cite{our_epl}, the interaction was not assumed to be additive, i.e., the strength of interaction of a growth cone with a fascicle did not depend on the number of axons in the fascicle.

{
Note that two fasciculated axons run parallel to each other with their separation in $x$-direction restricted within two
lattice spacings, the interaction range. Thus the typical width of a fascicle containing multiple axons remains 
2-3 lattice spacings in our model. We do not implement any on-site repulsion, i.e.,  axons are free to grow on top of each other.
}

In our model, we do not consider any relaxation dynamics of axon shafts.
This corresponds to the assumption of strong adhesion of axons  to the substrate, so that
the line tension on axons can not straighten out the local curvatures. 

\subsection{Turnover}
\label{turnover}
To capture the effect of neuronal turnover, each random walker is assigned 
a finite lifetime $\h$ from an exponential distribution of lifetimes
$\Pi(\h) d\h = \f{1}{T}\exp(-\h/T) d\h$ with mean 
$\la \h \ra = \int_0^\infty \h\, \Pi(\h) d\h = T$.
When the lifetime 
expires, the random walker and its entire trail (i.e., the whole axon)
is removed from the system. 
The mean number of axons in the system reaches the steady-state 
value $N(y)=N_0 \exp(-\be y)$, where $\beta = 1/T$ is the mean death rate per 
axon and the steady-state occupancy at $y=0$ is $N_0=\a/\be$.  
In the simulations, we use $T = 10^{5}$ time steps, and restrict 
ourselves to $y \le T/10$. The birth rate $\alpha$ is 
chosen so as to obtain the desired number of axons $N_0$, or equivalently, the 
desired axon density $\rho = N_0/L$ ($\r=1/2$ implies an average 
occupancy of one axon per site), where $L$ is the system size in 
$x$-direction. As we will show in detail, the time scales needed to achieve 
the steady state of fascicle size distribution can be very long compared to 
the time scale $T$ needed to achieve the steady state value of the total
number of axons.
A typical late-time configuration for a system of axons involving 
two distinct axonal types with type-specific interactions is shown in Fig.\ref{conf}$(b)$.

{
\subsection{Parameters}
\label{params}
In this section we briefly discuss the biological meaning and physical values of the parameters in our simulations.
In our model the interaction range in $x$- direction is chosen to be 2 lattice units. Since we assume only 
contact interactions, this range of 2 lattice units corresponds to the length of a filopodium which typically is
10 $\mu$m. Thus the lattice spacing in $x$-direction $\D x=5\, \mu$m. 

The  time step $\D t$ in our model needs to
be large enough to allow the growth cone to integrate a signal and react to it. 
 Ref.~\cite{Maskery2005} suggests this time scale to be of the order of  tens of seconds; 
 we choose $\D t = 60\,$s. This corresponds to a diffusion constant (in $x$-direction)
$(\D x)^2/2 \D t = 12.5\, \mu {\rm m}^2/$minute which compares well with {\em in vitro} observation 
for short time scales up to tens of minutes~\cite{katz1985}. 
Note that with this choice of $\D t$ the mean lifetime of an axon $T=10^5\, \D t$ used in our simulations corresponds to  
69.4 days, quite typical of axons of mouse olfactory sensory neurons~\cite{Nakatani03}. 

The lattice unit in $y$-direction is now chosen to give a reasonable growth velocity $v_y$. Choosing $\D y = 1\, \mu$m, 
we have $v_y = 60\, \mu$m/hour, which is a typical value for growing axons of sensory neurons~\cite{Honig98}.

Note that with the above mentioned choices, Fig.~\ref{conf}(b) corresponds to a system size of $800\, \D x = 4\,$mm in $x$- direction and
$10^4\, \D y = 10\,$mm in the $y$- direction. These dimensions are comparable to the size of the olfactory bulb in mice~\cite{Mombaerts}. 

The effective interaction energies $E_h$ and $E_o$ should be chosen to match the observed rates with which axons detach from fascicles. We introduce the quantity $\pi_d = \exp(E) / \D y$  that expresses  
the rate of detachment  of one axon per unit length of a two-axon fascicle given that the two axons interact via  $E=E_h$ or $E_o$. 
Thus a growth cone interacting with a fascicle of $n$ axons will follow the fascicle over the mean distance $L_y = \pi_d^{-n}$ before it detaches.
 It is not straigthforward to use published experimental images to deduce $E$ as usually, the size of the fascicle is not known, and the location 
 at which a growth cone first attached to the fascicle is not recorded.  The observed typical distance $L_y$ varies widely depending on the 
 specific neural system, ranging from tens of $\mu$m to centimeters. In our simulations, we use the range of homotypic interaction 
 strength $E_h = -4$ to -1, which corresponds to detachment rates $\pi_d = 0.02$  to 0.37 $\mu {\rm m}^{-1}$. In 
Table~\ref{parameters} we list the meaning  and values of  the parameters used in our model.

\begin{table}[htdp]
\caption{Parameters of the model}
\begin{center}
\begin{tabular}{|c|c|c|c|}
\hline
Symbol & Meaning  & Value & Value \\
  &  & (simulation) & (physical)\\
\hline
$\D x$ & Lattice spacing  & 1 & $5\,\mu$m \\
  & in $x$-direction &   &  \\
\hline
$\D y$ & Lattice spacing  & 1 & $1\,\mu$m \\
  & in $y$-direction &   &  \\
\hline
$\D t$ & Time step & 1 & $60\,$s \\
\hline
$T$ & Mean axonal lifetime & $10^5$ & $69.4\,$days\\
\hline
$N_0$ & Mean number of  & 50 to 200 & 50 to 200 \\
   &  axons at $y=0$ & &  \\
\hline
$L$ & System size  & 100 to 800 & 0.5 to 4 mm \\
   &  in $x$-direction & &  \\
\hline
$E_h$ & Homotypic  & $-4$ to $-1$ & detachment rate\\
            & interaction strength &     & $\pi_{dh}=0.02\,\mu {\rm m}^{-1}$\\
            &  &  & to $0.37\, \mu {\rm m}^{-1}$   \\
\hline
$E_o$ & Heterotypic  & $-0.1$ &  detachment rate \\
            & interaction strength &     & $\pi_{do} =0.9\, \mu {\rm m}^{-1}$ \\
\hline
\end{tabular}
\end{center}
\label{parameters}
\end{table}
}

\section{Overview}
\label{overview}
We show in this paper that simple directed growth of axons
that interact via a short-range attraction
leads to reliable formation of axon fascicles, 
in absence of any external chemical guidance cue.
Once a fascicle is formed its position does not move appreciably, however,
the fascicle size (number of axons present in the fascicle) fluctuates. 
The turnover of individual axons generates a slow dynamics of
reorganization of fascicles at a fixed $y$-level.

In the simplest case (Sec.~\ref{onesp}), the system contains only a single type of axons which  
grow and form fascicles via an energy-minimizing dynamics 
(strong inter-axon interaction), so that once attached the axons do not leave a fascicle. 
For this case one can uniquely assign a basin of each fascicle 
at any specified $y$-level (see Fig.~\ref{conf}($c$)). 
The basin size $D$ of a fascicle is the interval at the level $y=0$ between
the right-most and left-most axons belonging to the fascicle (Fig.~\ref{conf}($c$)).
Any axon growing from within this basin  must contribute to the fascicle size
unless it dies before reaching the specified $y$-level.
Thus the average number of axons that survives in a fascicle at 
level $y$ (in the steady state) is $\bar n = D\,\r \exp(-\be y)$.
The axons initiated at the opposite edges of the basin
are expected to meet each other in $y\simeq (D/2)^2$ steps
of random walk in $x$-direction. Therefore, one obtains the 
mean-field prediction for the mean fascicle size 
$\bar n (y) \simeq 2 y^{1/2}\, \r \exp(-\be y)$ up to $y \simeq (L/2)^2$,
where complete fasciculation occurs, i.e.,  $\bar n = N(y)$. 

In a system with finite detachment rates (Sec.~\ref{onespD}), however, growing axons can leave 
one fascicle and attach to another. Thus the fascicle basins overlap and
axons introduced in the basin of one fascicle can end up in a different fascicle.
Still the mean-field estimate of the increase of mean fascicle
size with $y$, shown above, turns out to remain approximately valid (Sec.~\ref{meanD}).
For higher detachment rates (weaker interactions), the prefactor of the 
power-law growth is reduced, corresponding to smaller fascicles.
In the limit of extremely weak interactions, each axon would grow
independently of the others, and no fasciculation is possible. 
This overall picture remains intact even for systems having multiple
axon types.
The dynamic steady state is characterized by a position ($y$) dependent 
distribution of fascicle sizes which shows a scaling law. The peak of
the distribution shifts towards larger fascicle sizes at higher $y$-levels (Sec.~\ref{steady} and Sec.~\ref{PsD}).

{
In a steady state configuration at fixed time $t$ (such as in Fig.~\ref{conf}$(c)$)  
the axon fasciculation with increasing $y$ 
may be formally viewed as the evolution of an one-dimensional diffusion-aggregation 
or diffusion-coalescence process~\cite{benAvraham}. 
These limited analogies can be used to approximately understand steady state properties
like distribution of fascicle sizes and inter-fascicle separation (Sec.~\ref{1dcomp}).
However, the full dynamics of our model has no simple one-dimensional
counter-part~\cite{our_epl}. 
A projection of the dynamics onto one dimension would involve
complicated long-time correlations between the random walkers.
}

The dynamics of fascicle reorganization can be characterized by 
the slow approach to steady state, or by the steady state
auto-correlation time for the mean fascicle size. These time scales grow 
with $y$ and may reach values orders of magnitude larger than $T$. 
In absence of detachment, the slowest mode of fascicle reorganization occurs via partial
exchange of neighboring basins (Sec.~\ref{eff2}). 
In the presence of detachment,  in addition, the basin of one fascicle
can easily drain to another. Thus the time scales decrease
with decreasing inter-axon attraction (Sec.\ref{meanD}).

In systems containing multiple types of axons with type-specific interactions,
we evaluate the degree of sorting $S$ that quantifies the type-wise 
purity of  the local environment  of  axons (Sec.~\ref{2sp}). 
At steady state, $S(y)$ shows a non-monotonic 
variation, with a maximum at some intermediate $y$. 
The position 
and the value of the maximum depend on system parameters like the 
mean density of axons $\r$ and the interaction strengths.
This non-monotonicity is due to the attractive heterotypic interaction
which merges mid-sized, relatively pure fascicles to form large impure fascicles.

\begin{figure}[t]
\psfrag{t/T}{$t/T$}
 \psfrag{ninf-n}{$n_\infty - \la \bar n \ra $}
 \psfrag{f(t)}{{$f(t)$}}
 \psfrag{y=103}{{$y=10^3$}}
 \psfrag{ y=102x}{{$y=10^2$}}
 \psfrag{ y=103x}{{$y=10^3$}}
 \psfrag{ 5000*x}{{$\r=1/8$}}
\includegraphics[width=8cm]{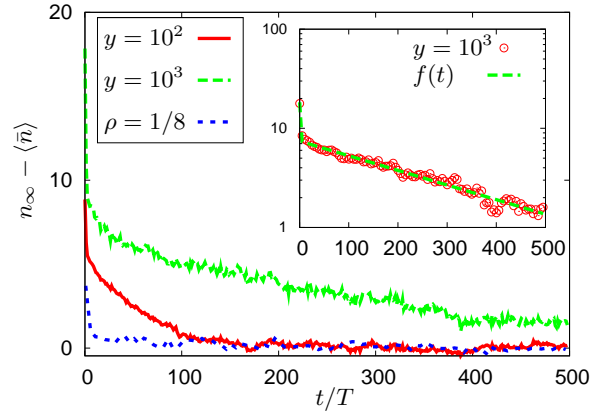}
\caption{(Color online)
Approach to the steady state in the $L=400$, $N_0=200$ system 
at representative $y$-levels indicated in the legend. 
The mean fascicle size $\la \bar n (t;y)\ra $ 
(averaged over $10^3$ realizations )
approaches $n_\infty(y)$ as $t\to\infty$. 
The data set labeled as $\r=1/8$ is from the $L=400$, $N_0=50$
system, collected at $y=10^4$.
Inset: Fitting of $n_\infty - \la \bar n (t,y)\ra $ to a function 
$f(t)=p \exp (-\be t) + q \exp (-t/\t_{ap})$ shown in a semi-log
plot. The data is the same as shown in the main figure at $y=10^3$. 
The fitting parameters are $n_\infty=42.94 \pm 0.07$, $p=13.98\pm 0.24$,
$q=7.43\pm 0.05$, and approach-to-steady-state time scale $\t_{ap}=(294 \pm 6)T$. 
{
We used the  Marquardt-Levenberg algorithm for nonlinear least-squares fitting as implemented in gnuplot version 4.4. 
}
}
\label{ap}
\end{figure}
\begin{figure}[t]
\psfrag{tap/T}{$\t_{ap}/T$}
 \psfrag{y}{$y$}
 \psfrag{ro=1/2    }{$\r=1/2$}
 \psfrag{ro=1/3    }{$\r=1/3$}
 \psfrag{ro=1/4    }{$\r=1/4$}
 \psfrag{ro=1/8    }{$\r=1/8$}
 \psfrag{tc, ro=1/2    }{$\t_c, \r={1}/{2}$}
 \psfrag{rty}{$y^{1/2}$}
\includegraphics[width=8cm]{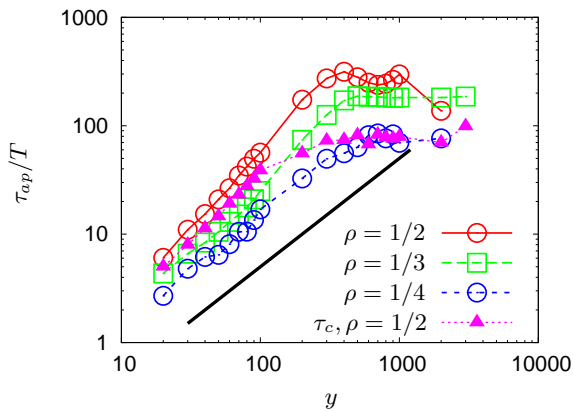} 
\caption{(Color online)
Approach-to-steady-state time scale $\t_{ap}$ as a function of 
$y$ at different axon densities. $\t_{ap}$ is extracted by
a fitting procedure as described for the inset of Fig.~\ref{ap}. 
The time series are collected over $t=500\, T$ and averaged over $10^3$ realizations . 
All the data were collected 
for a system of $N_0=50$ axons, 
with varied system sizes, 
(i) $L=100$ ($\r=1/2$), (ii) $L=150$ ($\r=1/3$), and (iii) $L=200$ ($\r=1/4$).
Correlation time $\t_c$: The last data set shows the steady state correlation time
$\t_c$ (in units of $T$) for a $L=100$, $N_0=50$ system calculated from the
correlation function $c(t)=\la\bar n(t) \bar n(0) \ra$.
$c(t)$ is evaluated by using the time series of $\bar n(t)$ collected between $t=200 T$ and
$2\times 10^4 T$, and averaged over $30$ realizations . 
A fitting of $c(t)=p+q \exp(-\be t)+ r \exp(-t/\t_c)$ allows us to extract $\t_c$  at different $y$-levels.
{
Fitting errors in $\t_{ap}$ and $\t_c$ are within $ 5\%$ (see the caption to Fig.~\ref{ap} (inset)
for the fitting  procedure and estimate of error in $\t_{ap}$).}
The thick solid line shows a power law $y^{2b}$ with $b= 1/2$.
}
\label{tap}
\end{figure}

\section{Single type of axons, no detachment} 
\label{onesp}
In this section we analyze the collective behavior of axons 
belonging to a single  type
following the energy-minimizing ``always attach, never detach" rule.
This model has been investigated in detail in a previous publication~\cite{our_epl}.
In this section, we extend the numerical and analytical results of Ref.~\cite{our_epl}. In the simulations, we use a modified implementation of the Monte Carlo update rules.
In contrast to Ref.~\cite{our_epl}, the strength of interaction between a
growth cone and a fascicle is assumed to be
proportional to the number of axons present
in the fascicle. Another difference is that we use parallel updates, instead of
sequential updates used in Ref.~\cite{our_epl}.
We therefore include a comparison to the main results we reported in Ref.~\cite{our_epl}, to show that these are not altered.

\subsection{Approach to steady state: mean fascicle size and time scale}

A typical late-time configuration for a system with $L=800$ and $N_0=100$
(density $\r=N_0/L=1/8$ at $y=0$)
is shown in Fig.~\ref{conf}$(c)$. 
With increasing $y$, the axons  
aggregate into a decreasing number 
of fascicles. 
The number of axons in 
a fascicle is referred to as the fascicle size $n$. 
At steady state, the mean 
fascicle size $\bar n$ at level $y$ may be estimated
as
$\bar n 
\simeq 2 \r y^{1/2} \exp(-\be y)$ 
up to $y \simeq (L/2)^2$, 
where complete fasciculation $\bar n=N(y)$ is expected~\cite{our_epl}.

\begin{figure}[t]
\psfrag{y}{$y$}
\psfrag{ns}{$n_s$}
 \psfrag{ro=1/2}{$\r=1/2$}
 \psfrag{ro=1/3}{$\r=1/3$}
 \psfrag{ro=1/4}{$\r=1/4$}
 \psfrag{ro=1/8}{$\r=1/8$}
 \psfrag{y^1/2}{{$y^{1/2}$}}
\includegraphics[width=8cm]{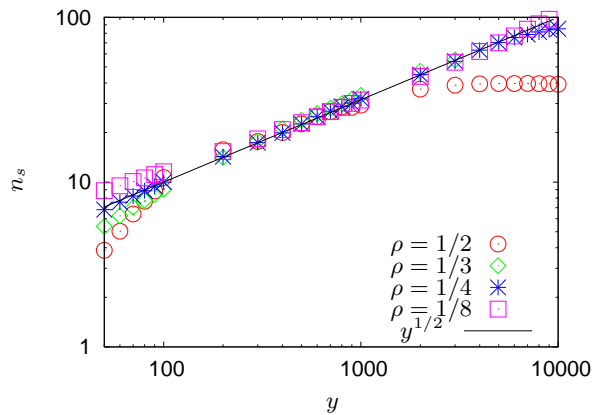}
\caption{(Color online)
Time-asymptotic fascicle size 
$n_\infty$ for systems with $N_0=50$ axons and system
sizes $L=100,\, 150,\, 200,\, 400$ 
(corresponding to $\r=1/2,\, 1/3,\, 1/4,\, 1/8$ respectively)
as a function of $y$. 
The time series are collected over $t=500\, T$ and averaged over $10^3$ realizations . 
The subtracted mean fascicle size $n_s=(n_\infty - c)\exp(\be y)/2\r$
corrected for the finite axon-lifetime  $T=1/\be$ 
is plotted as a function of $y$. The offset fascicle size $c$ is treated as a
fitting parameter, $c=8.17,\, 3.48,\, 1.56,\, 0.11$ for $L=100,\, 150,\, 200,\, 400$
respectively. 
{
Fitting errors in $n_s$ are within $3\%$.}
Data collected at the various densities collapse onto a power law 
$y^b$ with $b=1/2$. The largest system shows the widest power law regime.
}
\label{ninf}
\end{figure}
%
The measured mean fascicle size, obtained by averaging over all the 
existing fascicles at a given $y$  (Fig.~\ref{ap}), grows 
with time as $\bar n = n_\infty - p \exp (-\be t) - q \exp (-t/\t_{ap}) $, 
where  $\t_{ap}(y)$ defines the time scale of 
approach to the steady state value $n_\infty (y)$. 
The same behavior was observed earlier in simulations reported in Ref.~\cite{our_epl}. 
The semi-log plot in the inset of Fig.~\ref{ap} shows clearly the slow exponential
approach to the steady state mean fascicle size.
Using the above-mentioned double-exponential fitting we extract
the time scale $\t_{ap}$ and the steady-state mean fascicle
size $n_\infty$ at all the $y$-levels.

The approach-to-steady-state time scale 
$\t_{ap}$ increases with $y$.
$\t_{ap}$  can exceed the mean axon lifetime $T$ by orders of magnitude (Fig.~\ref{ap} and \ref{tap}). 
Note that $\t_{ap}$ is longer in a system with
larger density of axons $\r$ (Fig.~\ref{tap}). 
Ref.\cite{our_epl} discussed this point in detail. 
Further, asymptotically in $y$, we find $n_\infty = c+2\r y^b \exp(-\be y)$, with 
$b\approx 1/2$ (Fig.~\ref{ninf}) -- in good agreement with the 
mean-field prediction (Sec.~\ref{overview}). 

{\em Impact of interaction range:} To test the impact of the range of inter-axon interaction,
we have simulated a similar system with purely ``contact" interaction, i.e., the interaction range
is taken to be zero. 
With this reduction in the
range of interaction, 
we find that the emerging
time scales decrease.
For instance, for a system of $L=800$ and $N_0=100$ the approach-to-steady-state time
becomes $\t_{ap} \lesssim 10 T$.  
Thus an increase in the range of interaction increases the emerging time scales.
The steady state distribution of fascicle sizes shows the same scaling behavior as in the case
of nearest-neighbor interaction discussed in the following.

\subsection{Steady state}
\label{steady}
The steady state is characterized by the stationary distribution 
of fascicle sizes $P_s(n,y)$, defined as the number of fascicles of size $n$ 
at level $y$. 

\subsubsection{Steady state: scaling regime} 
\begin{figure}[t]
\begin{center}
\psfrag{Ps(n;y)}{$P_s(n,y)$}
 \psfrag{n}{$~~~~n$}
 \psfrag{APsny}{$AP_s(n,y)$}
\psfrag{Bn}{$Bn$}
\includegraphics[width=8cm]{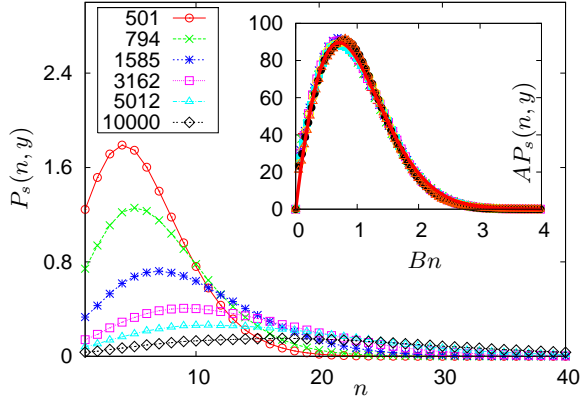}
\end{center}
\caption{(Color online)
Steady state distribution of fascicle sizes $P_s(n,y)$ 
(averaged over $10^4$ realizations  and the time interval
$10T\leq t\leq 25T$) for the $N_0=100$, $L=800$ system at $y$-levels 
indicated in the legend. 
Inset: A scaling with $B=1/\la n\ra$ and $A=\la n\ra^{2.1}$ collapses all data 
obtained for $y=1585,\, 1995,\, 3162,\, 5012,\, 6310,\, 7943,\, 10^4$ 
onto a single curve
$\phi(u)={\cal N} u \exp(-\nu u-\l u^2)$ 
with $u=n/\la n\ra$ and 
${\cal N}=274$, $\nu=0.78$, $\l=0.45$.  
}
\label{pkg-clps}
\end{figure}
For a system with $L=800$ and $N_0=100$, $P_s(n,y)$ is shown 
at a series of $y$-levels in Fig.~\ref{pkg-clps}. 
Within the range $y=10^3-10^4$ all data collapse onto a single curve after 
appropriate rescaling (Fig.~\ref{pkg-clps}). 
This data collapse implies the scaling law~\cite{our_epl}
\bea
P_s(n,y)= \la n(y)\ra ^{-r}\phi(n/\la n (y)\ra)
\label{scale}
\eea
with $r=2.1$ and the scaling function $\phi(u) = {\cal N} u \exp(-\nu u - \l u^2)$.
Note that the  steady state averaged fascicle size $\la n(y)\ra$ is
a quantity equivalent to the asymptotic $n_\infty(y)$ discussed in
the previous subsection.

{
The scaling law in Eq.~\ref{scale} can be justified starting from the 
assumption of homogeneity of fascicle size distribution 
$P_s(n,\l y)= \l^{-p} P_s(\l^{-q} n,y)$, 
with the exponents $p$ and $q$ undetermined at this stage. 
Noting that the mean
number of axons $N(y)=\int dn\, n\, P_s(n,y)$ and the mean number of fascicles 
$B(y) =\int dn  P_s(n,y)$, the homogeneity condition leads to the relations
$N(\l y) = \l^{-p + 2q} N(y)$, $B(\l y) = \l^{-p+q}B(y)$. Since by definition the 
mean fascicle size $\la n(y)\ra=N(y)/B(y)$, $\la n(\l y) \ra= \l^{q} \la n(y)\ra$. Invoking the 
mean-field prediction $\la n(\l y)\ra = \l^{b} \la n(y)\ra$ 
with $b=1/2$ (Sec.~\ref{overview} and \ref{onesp}.A), we find $q=b$.
 If $N(y)$ were independent of $y$, we would have had $p=2b$. 
 However, in fact $N(y)= N_0 \exp(-\be y)$. 
In the region $\be y<1$, we can write 
$p = 2b+\d$ with $\d \approx \be y/\ln y \ll 2b$.
Note that the relation $P_s(n,\l y)=\l^{-p}P_s(\l^{-b}n,y)$ can be recast in the form
$P_s(n,y) = \la n \ra^{-r}\phi(n/\la n\ra)$ where $r=p/b=2+\d/b \gtrsim 2$, in agreement with Eq.~\ref{scale}.
As $\d$ is $y$-dependent, the scaling of $P_s(n,y)$ is only approximate.

}

\subsubsection{Steady state: crossover to complete fasciculation}
\begin{figure}[t]
 \psfrag{Ps(n,y)}{$P_s(n,y)$}
 \psfrag{n}{$n$}
 \psfrag{y=500    }{$y=500$}
 \psfrag{y=1000    }{$y=1000$}
 \psfrag{y=2500    }{$y=2500$}
\includegraphics[width=8cm]{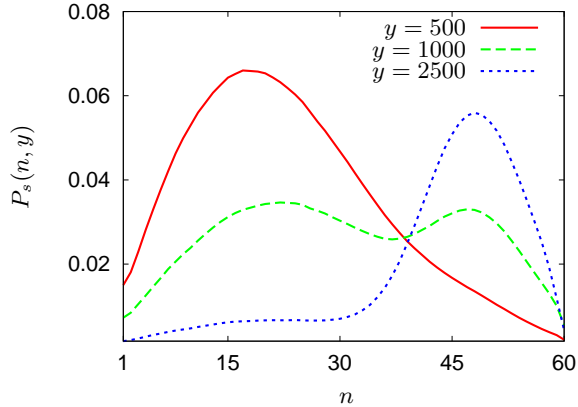}
\caption{(Color online)
Fascicle size distribution  at large $y$ for a system of
$L=100$ and $N_0=50$.
The data were averaged over $400T \leq t \leq 500T$ and 
$10^3$ realizations .
The single-peaked distribution characteristic of  the scaling
regime ($y = 500$)  crosses over to a distribution peaked near complete
fasciculation $n=N(y)=48.77$ at $y=2500~[=(L/2)^2]$ through a coexistence regime
showing a double maximum (at $y=1000$). 
}
\label{coexist}
\end{figure}
The steady state distribution changes its shape drastically beyond
the scaling regime. Near $y=(L/2)^2$, on an average all the
axons are expected to collapse onto a single fascicle, thereby
generating a distribution sharply peaked at $n=N(y)$. 
To demonstrate this fact we take a system of 
small size and high density $\r=1/2$ ($L=100$ and $N_0=50$). Within the scaling
regime ($y\lesssim 500$) the distribution function maintains the scaling form
$u \exp(-\nu u-\l u^2)$. However, at higher $y$-levels the 
distribution becomes bimodal with a new maximum appearing, characteristic 
of the complete fasciculation. 
This shows a coexistence of two preferred fascicle sizes.
Finally, at $y\simeq (L/2)^2$ the
whole weight of the distribution shifts to this new maximum and 
the distribution becomes  unimodal again (see Fig.~\ref{coexist}).
In systems with larger $L$, the regime of coexistence shifts towards higher
$y$-levels.

\subsubsection{Steady state: correlation time}
\label{time}
%

The dynamics in the steady state is characterized by the 
auto-correlation function for the mean fascicle size
$\bar n(t)$ at a fixed $y$-level: 
$c(t)=\la \bar n(t) \bar n(0)\ra$ which 
fits to the form $p+q\exp(-\be t)+r\exp(-t/\t_c)$ (as in Ref.\cite{our_epl}). 
The correlation time  $\t_c$ 
increases with $y$ and significantly exceeds the axon lifetime $T$.
We show this behavior  for systems at $\rho=1/2$ ($L=100$, $N_0=50$)
in Fig.~\ref{tap}. This shows a regime of approximate power law growth of the 
time scale $\t_c\sim y^{2b}$ with $b\approx 1/2$.

\subsection{Effective single-fascicle dynamics at fixed $y$}
\label{eff}
In this subsection, we review our analytical results from Ref.~\cite{our_epl}. 
The following subsection presents new results for 
the time scales arising from the interaction of two neighboring fascicles.

The concept of effective single-fascicle dynamics has been introduced in
Ref.\cite{our_epl}. 
The dynamics of a mean fascicle at level $y$ with $n(t)$ axons can be viewed 
as a stochastic process with gain rates $u_+(n)$
(for transitions $n \to n+1$)
and loss rates $u_-(n)$ (for transitions $n \to n-1$).
A fascicle loses axons only by the death of individuals,
therefore, $u_-(n)=\be n$~\cite{our_epl}. 

{
In absence of detachment events, 
any axon introduced within the basin (size $D$) of a fascicle 
(see Fig.~\ref{conf}(c)) can not escape the fascicle.
Moreover, some of the axons born in the neighboring inter-basin gaps (size $E$)
eventually join the fascicle under consideration. 
These two processes contribute to $u_+(n)$.
As was shown in
Ref.~\cite{our_epl}, the time series of $D(t)$ and $n(t)$ tend to co-vary. 
Thus treating the dynamics of 
$D$ as slave to $n$, we get a form $u_+ = a + b n$~\cite{our_epl}. Note that
the basin size $D$ can not exceed $2y$ or $L$, and $D>2 y^{1/2}$ occurs 
with low probability. Therefore a saturation of $u_+(n)$ is expected for 
large  values of $n$. 
The measured average gain and loss rates $u_\pm(n)$ 
obtained from our current simulations agree with the functional forms   
$u_+(n)=a_+ + b_+ n -c_+ n^2$ and
$u_-(n)=\be n$ (data not shown). 
The quadratic correction to linear growth captures the saturation of 
$u_+(n)$ at large $n$.}

The master equation of the growth-decay process for the effective 
single fascicle of size $n$ at level $y$ may be written as 
\bea
\dot P(n,t) &=& u_+(n-1)P(n-1,t)+ u_-(n+1) P(n+1,t)\nn\\ 
&-& [u_+(n)+ u_-(n)] P(n,t),
\label{master}
\eea
for $n>1$.
For the boundary state ($n=1$)  
$$\dot P(1,t) = J_+(y) + u_-(2) P(2,t)
                 - [u_+(1)+u_-(1)]P(1,t)$$
where $J_+(y)$ represents the rate with which new single 
axons appear between existing fascicles at $y$. 

The solution of the master equation at steady state was derived in Ref.~\cite{our_epl}
and has the form,
\bea
\be P_s(n,y) = J_+(y)~ n^\g \exp[-\ell (n-1) - \k (n-1)^2],
\label{lin2}
\eea
where $\g=a_+/\be-1$ and 
$\ell=1-b_+/\be$ and 
$\k=c_+/2 \be$.

From the master equation one can estimate the 
approach-to-steady-state time $\t_{ap}$ and the correlation time at
steady state $\t_c$~\cite{our_epl}. 
The correlation time $\t_c$ for the fascicle size $n$, 
near the macroscopic stationary point $n_s$ [$u_+(n_s)=u_-(n_s)$]
can be expressed~\cite{vanKampen} as
$\t_c=1/(u'_-(n_s)-u'_+(n_s)) =1/(\be - b_+ + 2c_+ n_s)$.
Under the linear approximation
of $u_+(n)=a_+ +b_+ n$ the approach-to-steady-state
time scale for the average fascicle size $\la n\ra$ 
can be written as $\t_{ap}=1/(\be - b_+)$ \cite{vanKampen}.

Further,
the mean lifetime of fascicles can be defined as 
$\t_f=[\int_1^\infty P_s(n,y)dn]/J_+(y)$ and is evaluated to obtain~\cite{our_epl}
\beq
\t_f=(T/2\k)\left[1-\left(\sqrt\pi e^{\f{\ell^2}{4\k}}(\ell-2\k) \text{erfc}(\ell/2\sqrt\k)\right)/2\sqrt\k\right]. \nn
\eeq

Notice that the above derivations of the time scales already
involved the numerical observation 
$u_+(n)=a_+ + b_+ n - c_+ n^2$.
Using the $y$-dependence of $b_+$ and $c_+$ obtained from numerical simulations, 
we found power law growth of the time scales, 
$\t_c \sim y^b$, $\t_{ap} \sim y^b$ and $\t_f \sim y^{2b}$ with $b \approx 1/2$~\cite{our_epl}.

In the following subsection, using a purely analytical, deterministic  treatment 
of the dynamics of two neighboring fascicles,  we  show that
a time scale growing as $y^{2b}$ emerges due to  
an exchange of basin size between the fascicles.


%
\begin{figure}[t]
 \psfrag{t}{$\t$}
 \psfrag{(a)}{$(a)$}
 \psfrag{di}{{$D_i$}}
 \psfrag{ei}{{$E_{i}$}}
  \psfrag{di1}{{$D_{i+1}$}}
 \psfrag{ei1}{{$E_{i+1}$}}
\includegraphics[width=7cm]{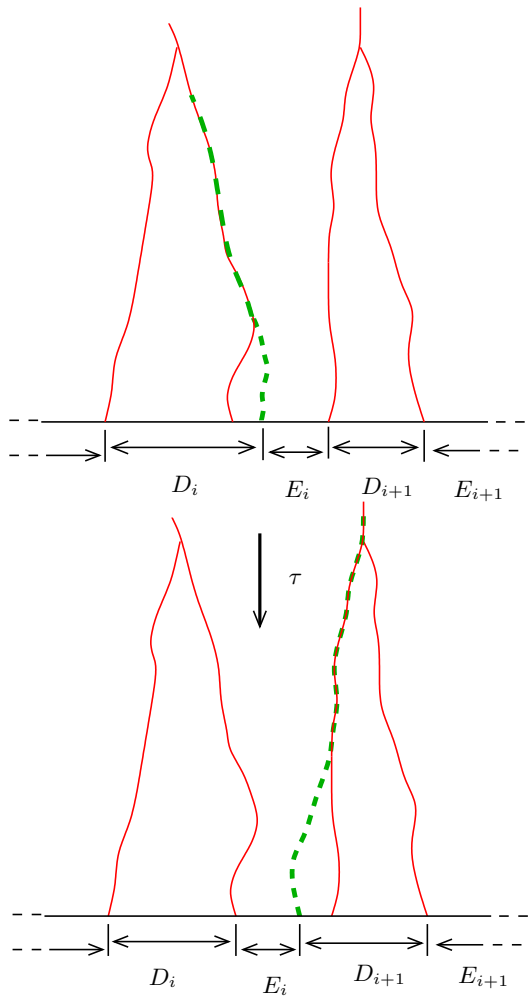}
\caption{(Color online) Illustration of basin size exchange, i.e., the slowest  mode in 
the effective dynamics of two fascicles (Sec.~\ref{eff2}). Two neighboring fascicles $i$ 
and $i+1$ are shown along with their corresponding basins $D_i$ and $D_{i+1}$,
and inter-basin gaps $E_i$ and $E_{i+1}$. 
The exchange of the boundary axon (dashed green line) between the two fascicles 
corresponds to a relaxation mode with 
time scale $\t \sim y^{2b}$ (see the main text). This increases the basin size $D_{i+1}$ 
at the cost of $D_{i}$, leaving the gap size $E_i$ unaltered.}
\label{basinDyn}
\end{figure}
\subsection{Effective dynamics of two interacting fascicles}
\label{eff2}
In this section we analytically explore the effective dynamics of
two neighboring fascicles. Under the ``always attach, never detach" 
rule the basins of neighboring fascicles do not overlap. 
Three dynamical variables characterize the dynamics of the fascicles: 
(i)~the number of axons $n_i$ present in the fascicle, 
(ii)~the basin size $D_{i}$ and 
(iii)~the separation $E_i$ between the $i$-th and $(i+1)$-th basin  (Fig.~\ref{conf}$(c)$). 
We explore the dynamics at a fixed $y$.
We consider the thermodynamic limit of large $N_0$ and $L$ (with a fixed density $\r=N_0/L$), 
and  express all the length scales in units of $L$ and number of axons
in units of $N_0$. Thus the reduced variables are $\n_i=n_i/N_0,~ \l_i=D_i/L$,
$\e_i = E_i/L$,  and the reciprocal system size $\tilde a=1/L$ 
has the meaning of a lower cut-off size in the continuum description. 
The effective equations of motion are
\bea
\f{d\n_i}{dt} &=& \be \left(\l_i +\f{\e_{i-1}+\e_{i}}{2} \right) - \be \n_i \nn \\
    \f{d\l_i}{dt} &=& \f{\a}{4}[\e_{i-1}(\e_{i-1}-\tilde a)+\e_{i}(\e_{i}-\tilde a)] 
                   - 2\be\d \f{\l_i}{\n_i-\d}\nn\\
    \f{d\e_i}{dt} &=& \be\d \left( \f{\l_i}{\n_i-\d} + \f{\l_{i+1}}{\n_{i+1}-\d} \right)
                  -\f{\a}{2}\e_i(\e_i-\tilde a).
\eea
where $\d = 1/N_0$.
First we describe the gain and loss terms in the dynamics of $\n_i$.
Axons born inside the basin of a fascicle contribute to the increase in $\n_i$,
hence the term $\be \l_i$ (we used  $N_0=\a/\be$ to express axon birth rate 
$\a$ in terms of $\be$).
A fascicle can lose axons only by individual axon deaths, thus
the loss term $\be \n_i$.
Any axon which is born in the inter-fascicle empty spaces $\e_{i-1}$  and $\e_i$
ends up in either of the two neighboring fascicles with
probability $1/2$, hence the gain  term $(1/2)\be (\e_{i-1} +\e_{i})$
\footnote{Notice that the fact that an axon born in a gap
may end up as a single axon 
has been ignored as the corresponding rate becomes negligible at high $y$.}. 

Next we consider the dynamics of basin size $\l_i$. 
A new axon can be born in the gap $\e_i$ with a rate $\a (\e_i -\tilde a)$ and attach
to the $i$-th fascicle with probability $1/2$. If it attaches it contributes  half the gap size 
$\e_i/2$ towards the basin size $\l_i$. Hence the gain term $(\a/4)\e_i(\e_i-\tilde a)$. 
A similar contribution to the gain in the basin size comes from the other neighboring gap $\e_{i-1}$.
The death of a boundary axon reduces the
basin size by an amount $\d \l_i/(\n_i - \d)$ (assuming no
double occupancy at a lattice point in the $y=0$ level).
The contributions of this loss coming from two
boundaries add up in the total loss term  $2 \times \be \d \l_i/(\n_i - \d)$.

Finally, we consider the dynamics of the inter-basin gaps $\e_i$. 
The death of boundary axons of neighboring fascicles $i$ and $i+1$ that 
border the $i$-th gap $\e_i$ contributes to the gain in the gap size.
Thus the gain terms $\be \d \l_i/(\n_i - \d)$ and $\be \d \l_{i+1}/(\n_{i+1} - \d)$.
Birth of an axon in the gap  reduces
the gap size under consideration. The rate of such an axon birth is $\a (\e_i - \tilde a)$
and on average this event reduces the gap size by an amount $\e_i/2$. Thus the loss term
$(\a/2) \e_i (\e_i - \tilde a)$.

We use a periodic boundary condition, such that the last fascicle is a nearest
neighbor of the first fascicle.
These equations obey the constraint of overall constant size $\sum_i (\l_i+\e_i)=1$.
It is important to note that the cut-off $\tilde a$ can be taken to zero 
meaningfully only after solving the differential equations.

For the simplest non-trivial case involving two fascicles, the
steady state that follows from these equations is characterized by
$\n_1=\n_2=1/2$, $\l_1=\l_2
=1/2-\tilde a/2-2\d-\tilde a\r/16$ and 
$\e_1=\e_2=\tilde a/2 + 2\d+\tilde a\r/16$. We perform a normal mode
analysis for small deviations from this steady state.
The constraint $\sum_{i=1,2} (\l_i+\e_i)=1$ implies that there are
only five independent deviations, $\d\l_1$, $\d\e_1$, $\d\e_2$, $\d \n_1$
and $\d \n_2$.
The linear stability analysis about the steady state shows that
all the five possible modes are stable. 
Among them, four modes are short-lived. For them the deviations decay
extremely fast with rates $\sim \be$. However, the fifth mode
which in the large-size limit can be written as
$$(\d n_1 = -1,\, \d D_1 \approx -1/\r,\, 
\d E_1 = 0,\, \d n_2=1,\, \d D_2\approx 1/\r)$$
takes a long time  to decay (see Fig.~\ref{basinDyn}).
It involves the loss of a boundary axon of one fascicle, which shrinks its
basin size by $\d D_1\approx -1/\r$, and simultaneously a gain of a
boundary axon for the other fascicle, increasing its basin size
by the equal and opposite amount $\d D_2 \approx1/\r$.
This operation leaves the inter-basin  gap unchanged ($\d E_1 = 0$)
and can be viewed as an exchange of basin space (Fig.~\ref{basinDyn}).
The deviations from steady
state in this mode decay over a very long time scale $\t\approx {\bar n}^2/3\be$
where $\bar n$ is the steady state value of the fascicle size  ($n_1=n_2\equiv \bar n$).

Using the  approximate growth of  mean fascicle size   $\bar n \sim y^b$ with $b=1/2$, 
we find a power law growth of this time scale  $\t\sim y^{2b}$.
Notice that the measured time scales $\t_{ap}$ and $\t_c$  obtained from MC simulations
show an increase with $y$ which approximately
obeys the power law $y^{2b}$ with $b=1/2$ (see Fig.\ref{tap}). 
At lower densities, the simulated data agrees better with the $y^{2b}$ 
power law. In the analytic calculation above, we
assumed single occupancy of the boundary sites of a fascicle basin
(removal of a boundary axon was assumed to reduce the basin size). 
At higher densities this assumption dose not hold, the boundary of
a basin does get multiply occupied by axons and thus we see
a departure 
from the $y^{2b}$ power law.

\section{Single type of axons, with detachment}
\label{onespD}
\begin{figure}[t]
\psfrag{ninf}{$n_s$}
 \psfrag{y}{$y$}
 \psfrag{(a)}{$(a)$}
 \psfrag{es=-4        }{{$E_h=-4$}}
 \psfrag{es=-2        }{{$E_h=-2$}}
  \psfrag{es=-1        }{{$E_h=-1$}}
 \psfrag{x^1/2        }{{$y^{1/2}$}}
\includegraphics[width=8.cm]{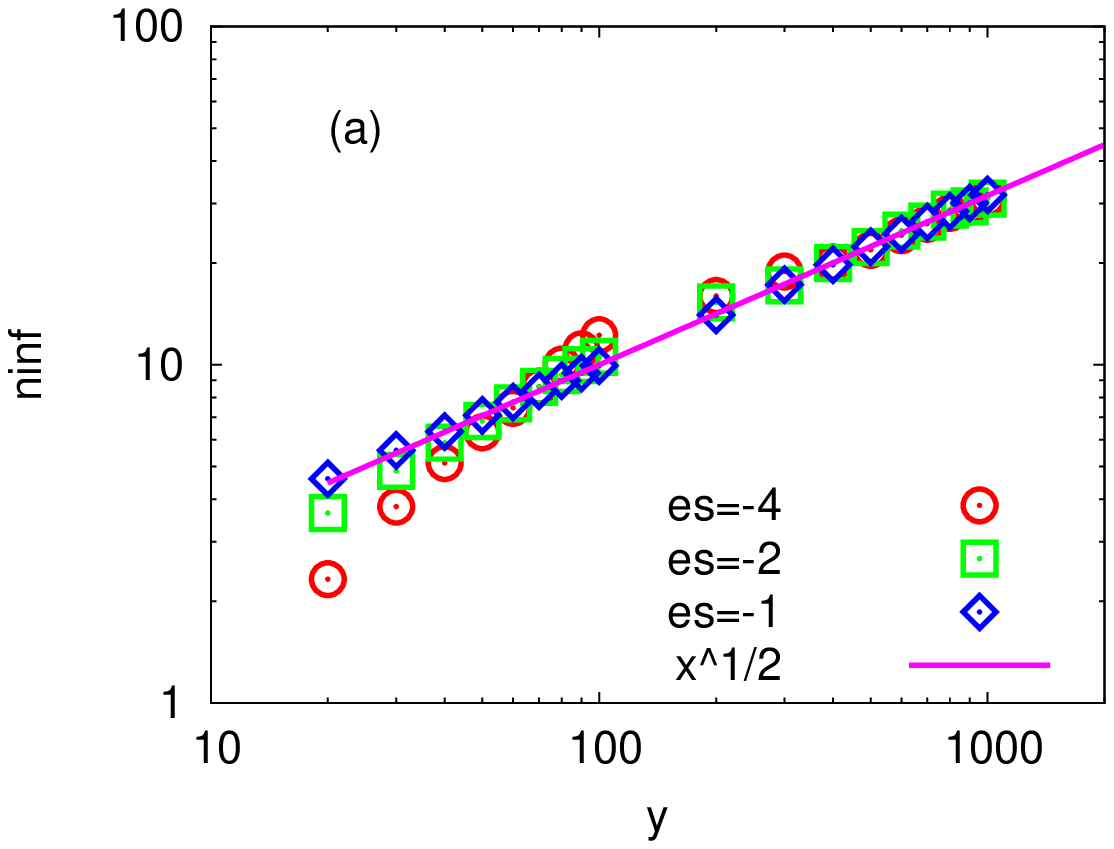}
\psfrag{tau}{$\t_{ap}/T$}
 \psfrag{y}{$y$}
 \psfrag{(b)}{$(b)$}
  \psfrag{es=-4        }{{$E_h=-4$}}
 \psfrag{es=-2        }{{$E_h=-2$}}
   \psfrag{es=-1        }{{$E_h=-1$}}
\includegraphics[width=8.cm]{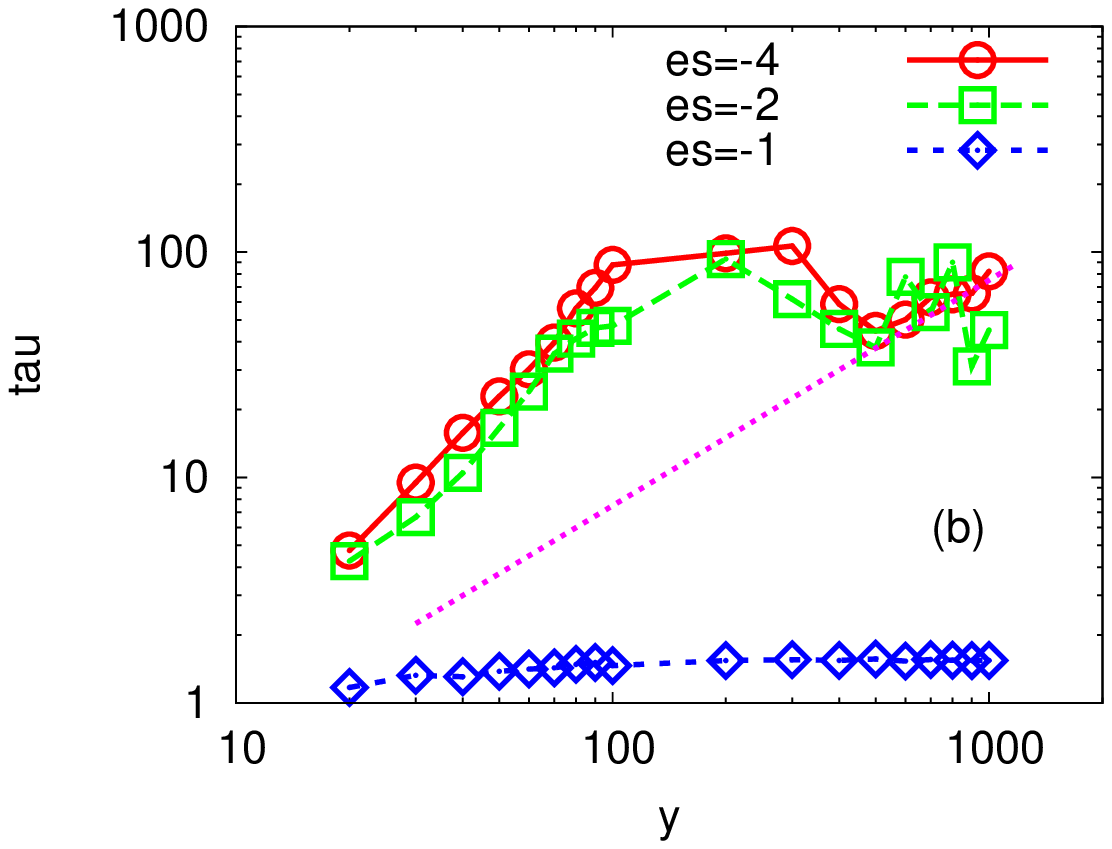}
\caption{(Color online) 
The time-asymptotic fascicle size and approach-to-steady-state 
time scale as a function of $y$ for a system with density
$\r=1/2$ ($N_0=100$ and $L=200$) at different inter-axon attractions $E_h$. 
The time series were collected over $t=500T$ and averaged over $10^3$ realizations .
($a$)~The subtracted time-asymptotic fascicle size $n_s=(n_\infty - c) \exp(\be y)/2 \r_{\rm{eff}}$ follows
a power law $y^{1/2}$. The effective density $\r_{\rm{eff}}$ and offset $c$ are treated as
fitting parameters with $c=5.44 \pm 0.62$, $\r_{\rm{eff}}=0.52 \pm 0.02$ for $E_h=-4$,
$c=3.44 \pm 0.26$, $\r_{\rm{eff}}=0.53 \pm 0.01$ for $E_h=-2$, and
$c=2.26 \pm 0.04$, $\r_{\rm{eff}}=0.41 \pm 0.002$ for $E_h=-1$.
($b$)~The approach-to-steady-state time scale  $\t_{ap}$ gets smaller for weaker 
attractions $E_h$, however, shows the initial power law growth unless $E_h\gtrsim -1$.
The dotted line shows a power law $y^{2b}$ with $b=1/2$.
{
Fitting errors in $\t_{ap}$ are within $ 5 \%$.}}
\label{timePd}
\end{figure}
\begin{figure*}[t]
 \psfrag{Ps(n;y)}{$P_s(n,y)$}
 \psfrag{n}{$n$}
 \psfrag{APsny}{$AP_s(n,y)$}
\psfrag{Bn}{$Bn$}
 \psfrag{(a)}{$(a)$}
 \psfrag{y=10       }{{$y=10~~~~~~$}}
 \psfrag{y=10^2       }{{$y=10^2~~~~$}}
 \psfrag{y=10^3       }{{$y=10^3~~~~$}}
 \psfrag{y=10^4       }{{$y=10^4~~~~$}}
\includegraphics[width=8.2cm]{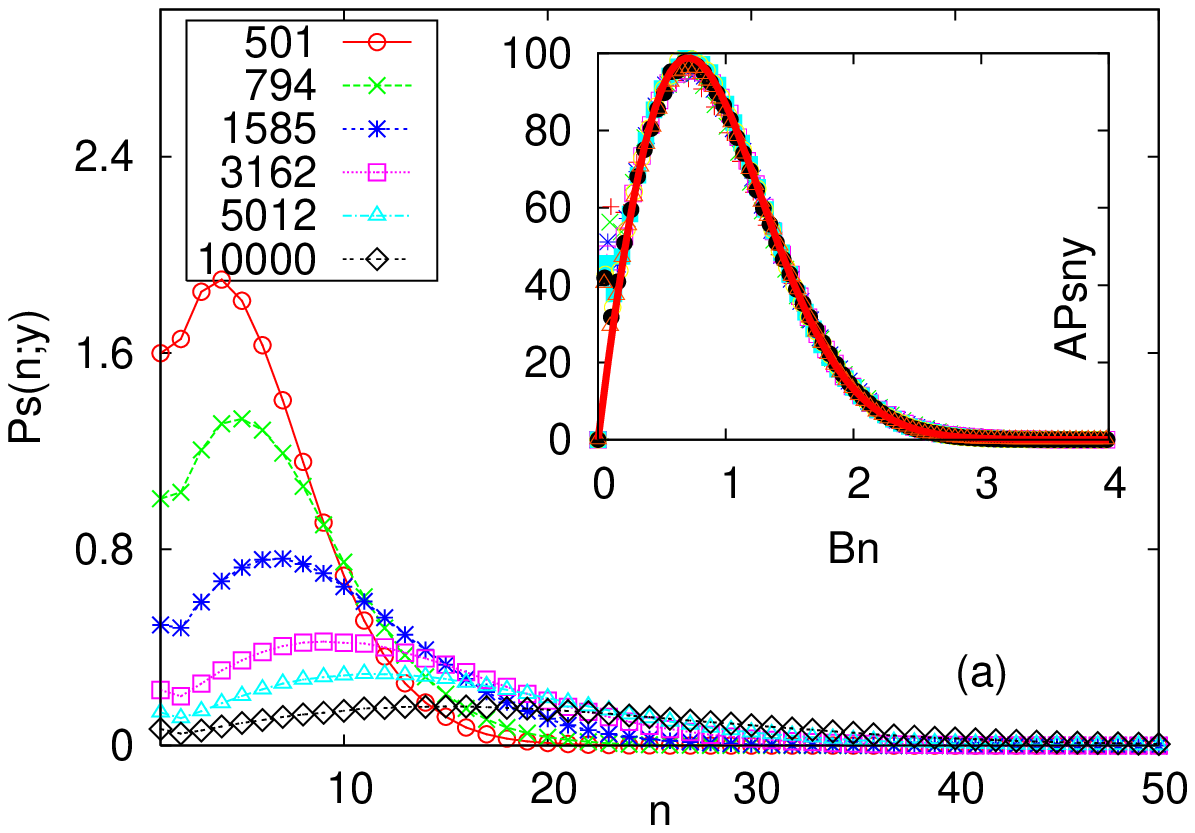}
\psfrag{Ps(n;y)}{$P_s(n,y)$}
 \psfrag{n}{$n$}
 \psfrag{APsny}{$AP_s(n,y)$}
\psfrag{Bn}{$Bn$}
 \psfrag{(b)}{$(b)$}
 \psfrag{ro=1/8       }{{$\r=1/8~~~~$}}
 \psfrag{ro=1/2       }{{$\r=1/2~~~~$}}
\includegraphics[width=8.2cm]{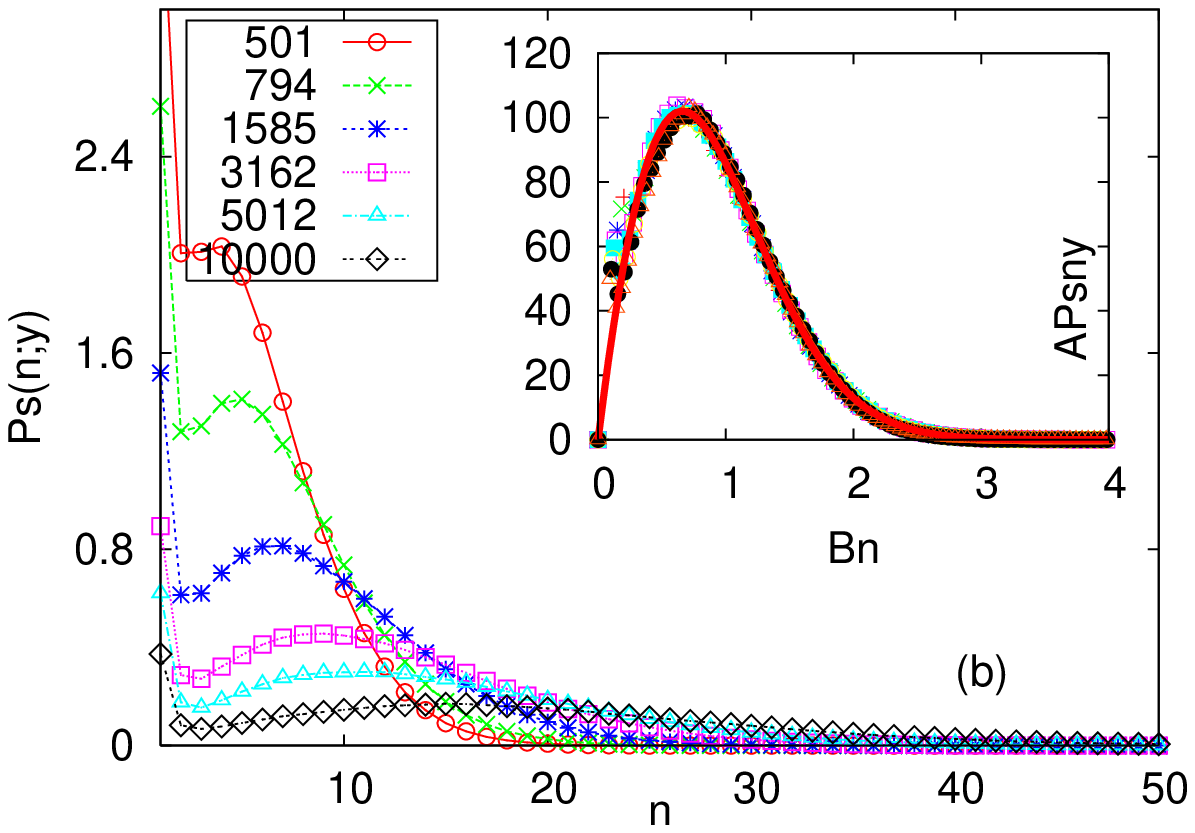}
\psfrag{Ps(n;y)}{$P_s(n,y)$}
 \psfrag{n}{$n$}
 \psfrag{APsny}{$AP_s(n,y)$}
\psfrag{Bn}{$Bn$}
 \psfrag{(c)}{$(c)$}
 \psfrag{ro=1/8    }{{$\r=1/8~~~~$}}
 \psfrag{ro=1/2    }{{$\r=1/2~~~~$}}
\includegraphics[width=8.2cm]{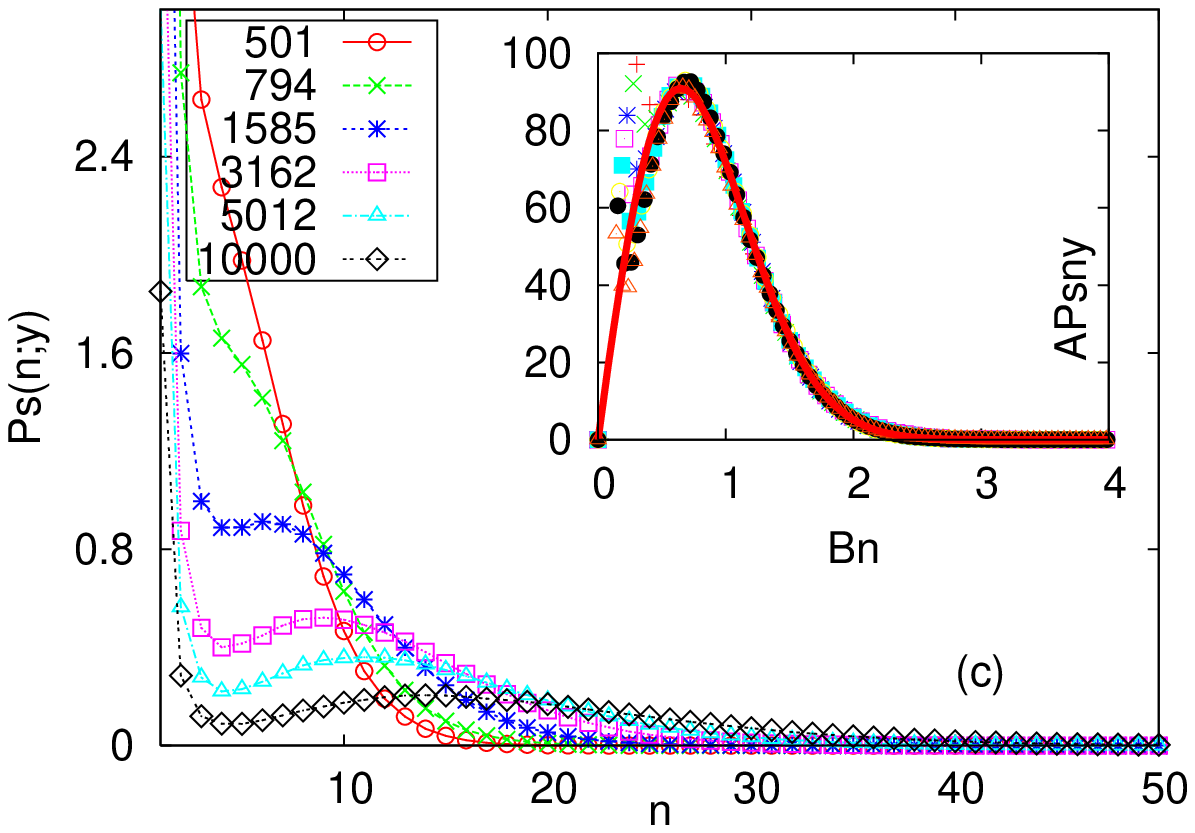}
\psfrag{Ps(n,y)}{$P_s(n,y)$}
 \psfrag{n}{$n$}
 \psfrag{Psny}{$P_s(n,y)$}
\psfrag{Bn}{$Bn$}
 \psfrag{(d)}{$(d)$}
 \psfrag{t_s,ro=1/8    }{{$\r=1/8,\t_s~~~~$}}
 \psfrag{t_n,ro=1/8    }{{$\r=1/8,\t_n~~~~$}}
 \psfrag{t_s,ro=1/2    }{{$\r=1/2,\t_s~~~~$}}
 \psfrag{t_n,ro=1/2    }{{$\r=1/2,\t_n~~~~$}}
\includegraphics[width=8.2cm]{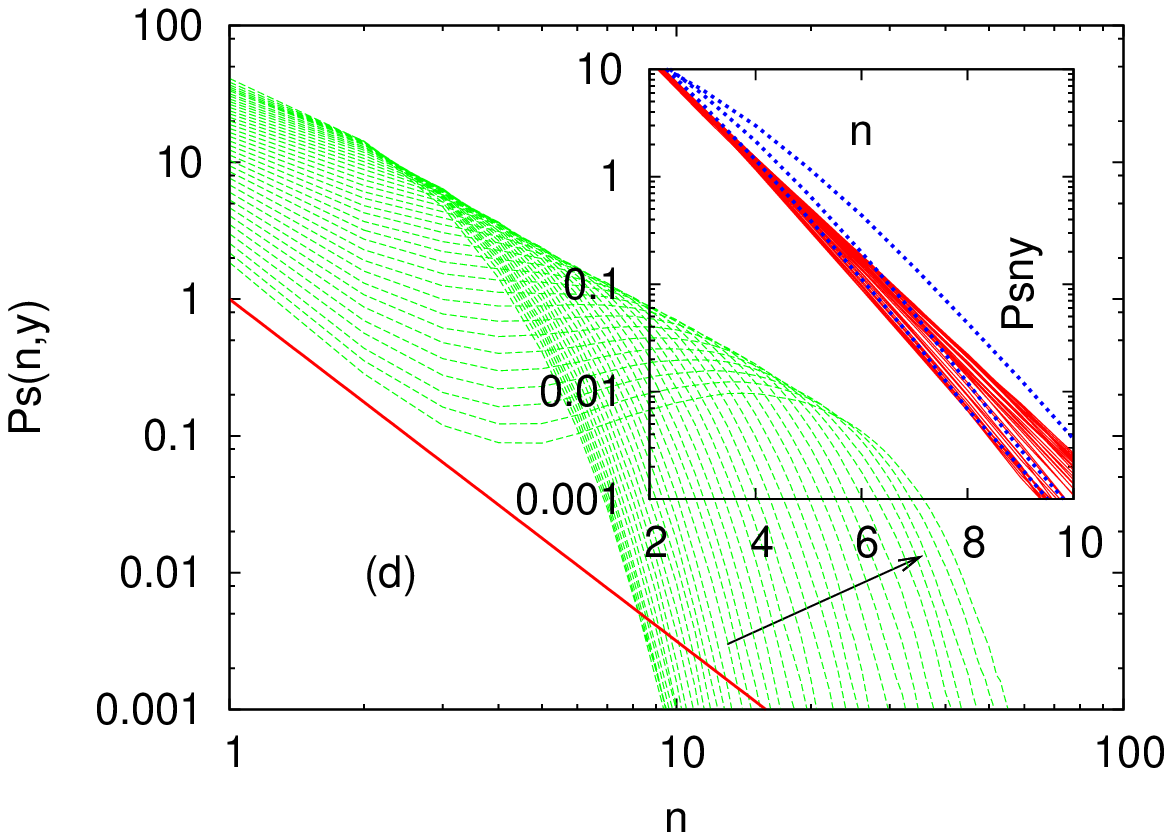} 
\caption{(Color online) Steady state distribution of fascicle sizes 
$P_s(n,y)$ (averaged over $10^4$ realizations  and the time interval
$10T \leq t\leq 25T$) for the $N_0=100$, $L=800$ system at different strengths of inter-axon
interactions
($a$)~$E_h=-4$,
($b$)~$E_h=-2$, and
($c$)~$E_h=-1$. 
The insets show the data collapse. 
A scaling with $B=1/\la n\ra$ and $A=\la n\ra^{r}$ collapses  data 
obtained for $y=1585,\, 1995,\, 3162,\, 5012,\, 6310,\, 7943,\, 10^4$ 
onto a single curve
$\phi(u)={\cal N} u \exp(-\nu u-\l u^2)$ 
with $u=n/\la n\ra$ and 
($a$)~${\cal N}=262.6$, $\nu=0.37$, $\l=0.74$, $r=2.1$
($b$)~${\cal N}=318$, $\nu=0.67$, $\l=0.64$, $r=2.1$ and
($c$)~${\cal N}=244.5$, $\nu=0.18$, $\l=1.05$, $r=2$. The fitting to obtain the scaling
functions is  done above $u=0.4$.
($d$)~ The same distributions as in ($c$) presented in a log-log plot. 
$P_s(n,y)$ decreases with $n$ in the small $n$ regime.  
{
The  thick line (red) denotes a power law $n^{-5/2}$.}
The arrow denotes the direction of increasing $y$ ($y=1,\dots,10^4$).
Note that the curve obtained at $y=10^4$ is the closest to the $n^{-5/2}$
line at small $n$.
For $y>500$, the initial decrease of $P_s(n,y)$ with $n$ is followed by a 
subsequent increase which merges with the scaling function. 
At $y \leq 20$, the tail of the distribution is exponential 
(non-interacting limit). 
Inset of ($d$): The solid (red) lines are the fascicle
size distributions for $E_h=-0.1$ at all $y$-levels ($10< y \leq 10^4$). 
It shows a clear single exponential decay
characterizing the (almost) non-interacting axons. 
This should be compared with the dotted (blue) lines that show approximate
exponential decay of the distributions in the range $y\leq 20$ for $E_h=-1$.}
\label{1spT1}
\end{figure*}
In the previous section, we concentrated on  the ``zero-temperature",  energy-minimizing dynamics, in which axons cannot detach from a fascicle once they become part of it. Now we extend our analysis to ``finite-temperature"  Monte-Carlo dynamics, in which the detachment of growth cones from fascicles become possible. It is important to analyze this general case as in the experimental studies of fasciculation dynamics~\cite{Honig98, Wolman07,Lin94}, defasciculation events are clearly observed. An additional reason is that the detachment from fascicles is crucial for the formation of pure fascicles in a system containing multiple axon types -- see Section~\ref{2sp}.

As described in Sec.~\ref{model}, 
in our Monte Carlo simulations  (with effective temperature set to unity), 
a randomly attempted move to
left (right) is accepted with probability $p_L =$ min$[1,\exp(-\d E_l)]$ 
($p_R =$ min$[1,\exp(-\d E_r)]$) where $\d E_l$ and $\d E_r$ are
evaluated based on the axon occupancy numbers and on the additive
interaction energy per axon $E_h$ ($<0$). 
The detachment rate of a growth-cone following a fascicle of size $n$
is $\exp(n E_h)$.
Note that for weaker $E_h$, the detachment rate is larger.

\subsection{Impact of detachment on $n_\infty$ and $\t_{ap}$}
\label{meanD}
The approach-to-steady-state time scale $\t_{ap}$ decays 
with increasing detachment rates (decreasing inter-axon attraction $|E_h|$).
We perform the approach to steady state data analysis in the same
manner as we did for  the purely energy-minimizing
dynamics discussed in the previous section.
This analysis gives an estimate of the time-asymptotic mean fascicle size
$n_\infty$ as well as the approach-to-steady-state 
time scale $\t_{ap}$ as a function of $y$. We perform this analysis
for a system of $N_0=100$ and $L=200$ at various strengths of 
inter-axon interaction $E_h$. 
Strictly speaking, the presence of detachment invalidates our earlier mean field argument for the growth of fascicle size with $y$. While the concept of the basin of a fascicle is still meaningful, the basin can ``leak", i.e.,  newly growing axons can escape from it through the detachment process. Despite this, for most interaction strengths our numerical results obey 
$n_\infty = c + 2 \r_{\rm{eff}} y^{1/2} \exp(-\be y)$ with the 
fitting parameter $\r_{\rm{eff}}$ taking the place 
of $\r$ [Fig.~\ref{timePd}($a$)].
{
The value of $\r_{\rm{eff}}$ is the 
smallest for the weakest attractive
interaction plotted in Fig.~\ref{timePd}$(a)$.}
For extremely weak attraction (e.g., $E_h=-0.1$) 
almost no fasciculation can occur {
($\r_{\rm{eff}} \simeq 0$)} and
therefore $n_\infty$ remains independent of $y$ (data not shown).

As shown in Fig.~\ref{timePd}($b$), the time scale $\t_{ap}$  for the approach to steady state decays with reduced inter-axon attraction (i.e., with increased detachment rate). When $E_h \leq - 2$, the growth of $\t_{ap}$ with $y$ is similar to the one we observed for the ``zero-temperature" dynamics. 
At $E_h=-1$, however, we find that 
$\t_{ap}$ becomes comparable to the mean axonal lifetime $T$,
even though at this value of $E_h$ the power-law growth of the mean fascicle size with $y$ is
still maintained. 

The reduction of $\t_{ap}$ with increased detachment rate may be understood as a result 
of enhanced interaction between neighboring fascicles. As we noted before, the slowest time scale
for the case of purely energy-minimizing dynamics is due to the 
very slow process of exchange of basin size between  neighboring fascicles. In presence of finite
detachment rates, axons from one fascicle can detach and connect to a neighboring fascicle,
thereby opening up a new and faster mode of interaction between neighboring fascicles.

 \subsection{Impact of detachment on the fascicle size distribution}
 \label{PsD}
In this subsection we show that even in presence of an appreciable detachment
rate ($E_h\leq -1$), the scaling of fascicle size distribution  persists and the scaling
function retains its overall functional form. However, with increasing detachment
rates (decreasing $|E_h|$), the
small $n$ portion of the fascicle size distribution gains at the
cost of bigger fascicles and deviates from the scaling form.  

Fig.~\ref{1spT1} shows the
steady-state fascicle size distributions for a system of $L=800$ and
$N_0=100$.
As the attraction is decreased from
$E_h=-4$ to $E_h=-1$, we observe an overall increase in the
small $n$ portion of the distribution.
Except for the lowest values of $n$, we obtain
data collapse implying the scaling form
$$
P_s(n,y)=\la n\ra^{-r}\phi(n/\la n\ra)
$$
where $r=2.1$ for $E_h=-4,\,-2$, and $r=2$ for
$E_h=-1$. 
Similarly to the case of strictly energy-minimizing dynamics, 
this scaling is observed only in the intermediate range of $y$ values,  $10^3 <y< 10^4 $, and the scaling function 
is of the form $\phi(u)={\cal N} u\exp(-\nu u -\l u^2)$ with $u=n/\la n\ra$.

The small-$n$ part of the distribution does not scale. 
$P_s(n,y)$ is large at $n=1$, and  drops to lower values with increasing $n$, 
before it increases again to follow the scaling function.
The functional form of the initial decay of $P_s(n,y)$ with $n$ depends on $y$ 
and also $E_h$, as we show in detail in Fig.~\ref{1spT1}$(d)$.

The discussion in this section shows that there are parameter regimes,
e.g., the $E_h=-1$ case discussed above, where the emergent time scales are
comparable to the mean lifetime $T$ of single axons, and at the same time
the steady state statistics (the mean fascicle size and the fascicle 
size distribution) obey the overall features demonstrated by the 
energy-minimizing dynamics. This parameter regime might be utilized to attain 
a fasciculation pattern of this type in a relatively short time.

\begin{figure*}[t]
\begin{center}
 \psfrag{AH(dx)}{{\small$A\, H(\Delta x)$}}
\psfrag{Bdx}{$B\, \Delta x$}
   \psfrag{(a)}{$(a)$}
\includegraphics[width=7cm]{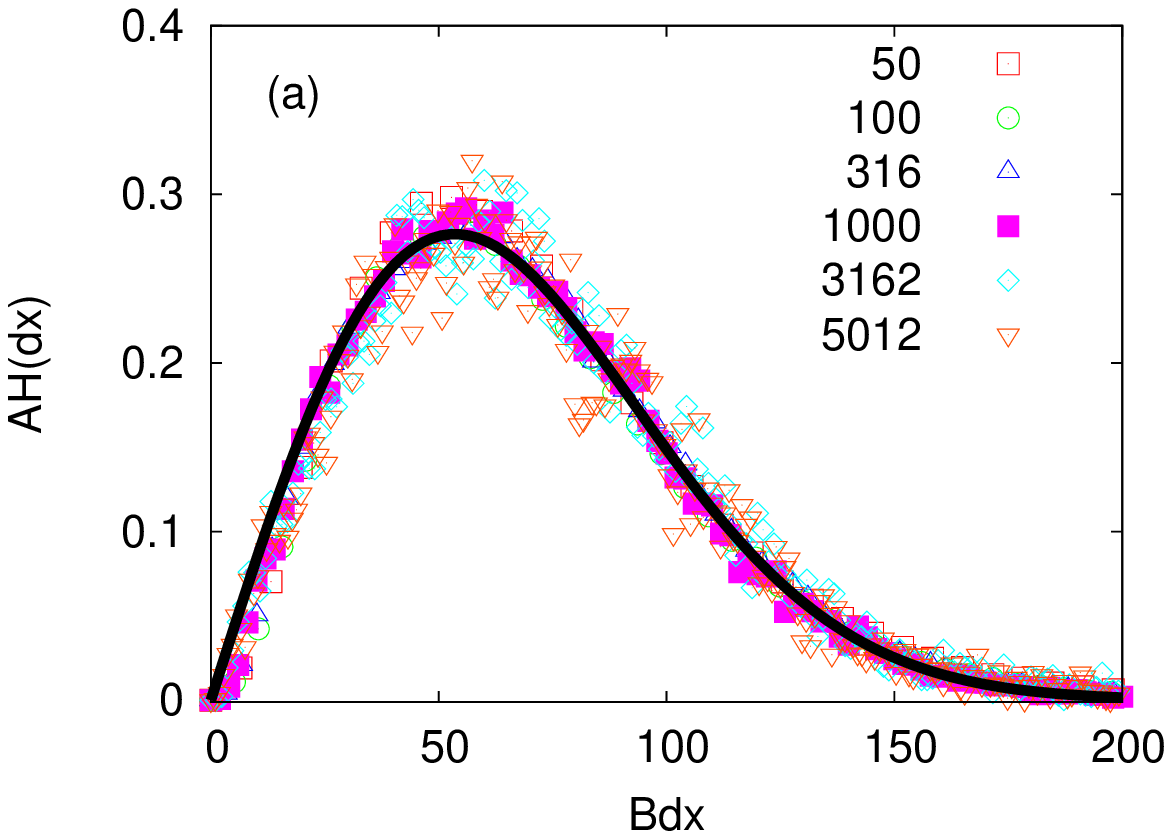}
\psfrag{A}{$A$}
 \psfrag{B}{$B$}
  \psfrag{A,B}{$A,\,B$}
   \psfrag{y}{$y$}
      \psfrag{(b)}{$(b)$}
\includegraphics[width=7cm]{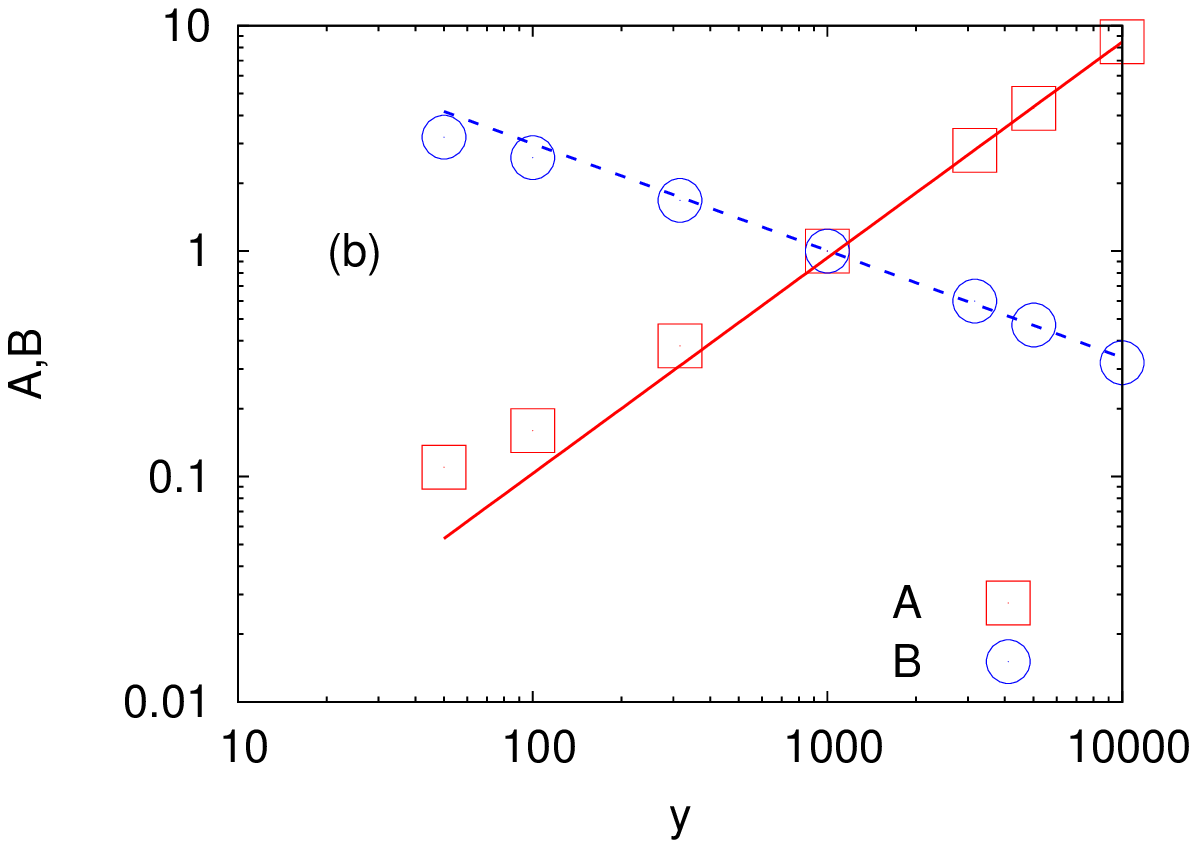}
\end{center}
\caption{(Color online) Histogram of inter-fascicle separation $\Delta x$.
($a$)~Rescaling of the histogram $H(\Delta x)$, obtained at various
$y$-levels indicated in the legend, leads to data collapse. The line through the collapsed data
is a function ${\cal P} \Delta x \exp(-{\cal Q} \Delta x^2)$ with ${\cal P} =0.0085$ and ${\cal Q}= 0.0002$.
$(b)$~The scale factors $A,\,B$ obtained at different $y$-levels show power-law dependence
$A \sim y^{0.96}$ and $B\sim y^{-0.47}$ indicated by the lines through the data points.   
}
\label{gapDist}
\end{figure*}

{
\section{Relation to particle aggregation and coalescence in one dimension}
\label{1dcomp}
As we pointed out in Sec.~\ref{overview}, in a steady-state configuration at fixed time $t$,  the axon fasciculation with increasing $y$  
may be formally viewed as the evolution of a one-dimensional (1d) reaction-diffusion process, where the $y$ coordinate 
takes the meaning of time. As we show in this section, this limited analogy can be used to 
approximately understand some steady state properties,  e.g., the distributions of fascicle sizes and of 
inter-fascicle spatial separations.  
We stress, however, that  the full dynamics of our system can not be mapped on to a 1d reaction-diffusion system. 
The process of axon turnover, which is crucial for the dynamical properties of our system, does not have any analog in the 1d models we discuss in this section. 

\subsection{In absence of detachment}
In this subsection we discuss the relation of 
our basic model, in which axons cannot detach from fascicles,
to irreversible aggregation and coalescence processes.

Interpreting the fasciculation of axons with increasing $y$ (in a steady state
configuration)  as 1d {\em irreversible aggregation} of particles 
($m A + n A \to (m+n)A$)~\cite{redner,benAvraham},
we find a prediction $u\exp(-\l u^2)$ (see equation 8.4.24 in ~\cite{redner})  for the fascicle-size distribution, 
which is  similar to the true distribution $u\exp(-\nu u-\l u^2)$ (Sec.~\ref{steady})~\cite{our_epl}. 
Note that the distribution obtained from the mapping to irreversible aggregation lacks 
the exponential part $\exp(-\nu u)$. As we explained in Ref.~\cite{our_epl}, having $\nu=0$ in our model would require $\tau_{\rm ap}=\infty$. The absence of the exponential part is therefore consistent with the absence of turnover-based dynamics in the 1d analogy.

The merging of fascicles  with growing $y$ 
may alternatively be interpreted as an {\em irreversible coalescence} process 
$A+A \to A$~\cite{Avraham1990}, viewing each fascicle as a particle $A$. 
The pattern formation in 1d irreversible coalescence  had 
been quantified by the inter-particle distribution function (IPDF)~\cite{Avraham1990},  
the distribution of distance between neighboring particles. 
The steady state of this process is trivial, a completely empty 
space. The IPDF is obtained at finite time $t$ before this steady state arrives, and has 
the form $(x/4{\cal D}t)\exp(-x^2/8{\cal D}t)$ where $\cal D$ denotes the particle diffusion constant~\cite{Avraham1990}.
The change in the inter-fascicle separation distribution with increasing $y$ in our model 
may be viewed as equivalent to the {time} evolution of  IPDF
in irreversible coalescence. This leads us to a prediction of 
the  distribution of inter-fascicle separation $H(\Delta x,y) \approx (\Delta x/y)\, \exp(-\mu\, \Delta x^2/y)$.
Using the same stochastic simulation that we used to obtain Fig.~\ref{pkg-clps}, we calculated the histogram of spatial separations 
$\Delta x$ between fascicles identified at various $y$-levels. 
Note that the separation between two neighboring fascicles  $\Delta x$ is measured at the same $y$-level at which the 
fascicles are identified, and is different from the gap between fascicle basins $E$  (shown in Fig.~\ref{conf}).
A rescaling of $H(\Delta x)$ by a factor $A(y)$ and $\Delta x$ by $B(y)$ 
leads to a data collapse (Fig.~\ref{gapDist}(a)), and approximate power law dependences 
$A \sim y^{p}$ and $B\sim y^{-q}$ with $p=0.96$ and $q=0.47$ (Fig.~\ref{gapDist}(b)). 
This result is in reasonable agreement with the above-mentioned form of  $H(\Delta x,y)$  
that predicts a scaling function  ${\cal P} \Delta x \exp(-{\cal Q} \Delta x^2)$ (Fig.~\ref{gapDist}(a)),
and scaling exponents $p=1$ and $q=1/2$.

\subsection{In presence of detachment}
In this subsection we discuss analogies of 
our model of axon fasciculation in presence of detachment 
to reversible aggregation and coalescence processes in 1d. Due to specific features of our model, only qualitative analogies to models from the 1d literature can be made.

Detachment events are partially captured when
the axon fasciculation with increasing $y$ in a fixed-time configuration 
is formally viewed as a 1d diffusion with {\em reversible aggregation},  
the chipping model: $m A + n A \rightarrow (m+n) A$
and $m A \rightarrow (m-1)A + A$~\cite{mustansir}, denoting each axon by a particle $A$. 
(Note, however, that steady state configurations of our model show splitting of fascicles with increasing $y$ (e.g., see Fig.~\ref{conf}(b)) 
into two fascicles containing multiple axons. The chipping model does not include the analog of such a process).
The chipping model posseses a non-trivial steady state ($t \rightarrow \infty$, corresponding to $y \rightarrow \infty$ within our model). It shows a dynamic phase transition associated with particle density $\rho$~\cite{mustansir,Rajesh2001}.
The steady state distribution of clusters of size $n$ is predicted to be $P_s(n)\sim \exp(-n/n^\ast)$ at 
$\rho <\rho_c = \sqrt{1+w}-1$,
where $w$ denotes a constant single particle chipping rate. At the critical density $\rho=\rho_c$, the distribution 
changes its shape to $P_s(n) \sim n^{-5/2}$.  At density above $\rho_c$ this power-law distribution remains unaltered, and in addition to the power law distributed clusters one gets a single cluster of diverging size~\cite{mustansir}. 
Fig.~\ref{1spT1} shows fascicle size distributions obtained from our model at various interaction strengths $E_h$. Note that, in contrast to the chipping model, in our model the rate $w = 1/[1+ \exp(-n E_h)]$, with which a single axon detaches from a fascicle, depends on the fascicle size $n$. 
Only for the weakest interaction $E_h=-0.1$ (inset of Fig.~\ref{1spT1}(d)), 
the detachment rate from a two-axon fascicle $w=0.45$ corresponds to a critical density $\rho_c=0.2$  which is greater 
than the axon density $\rho=1/8$. The corresponding fascicle size distribution shows a form consistent with
$\sim \exp(-n/n^\ast)$ (inset of Fig.~\ref{1spT1}(d)).
For $E_h=-1$ the detachment rate from a two-axon fascicle $w=0.12$ corresponds to a critical density 
$\rho_c=0.06 < \rho$ ($= 1/8$). The fascicle size distribution in the region of small $n$, for $E_h=-1$,  shows rough
agreement with the power law  $\sim n^{-5/2}$ (Fig.~\ref{1spT1}(d)). 
This change in shape of the fascicle size distribution from an exponential decay to a power-law decay
at small $n$, thus, is consistent with the dynamical phase transition predicted by the chipping model.
Since in our model the detachment rate of axons exponentially decays
with fascicle size $n$, for larger fascicles 
the detachment rate $w$ gets so small that the analogy with the {\em reversible}  chipping model  breaks down, and the 
behavior of the system becomes analogous to {\em irreversible} aggregation.
The fascicle size distribution in the region of larger $n$ (Fig.~\ref{1spT1})  
becomes indistinguishable from axon fasciculation in absence of detachment. 

%

We note that, for our model in presence of detachment, 
the change in the  distribution of inter-fascicle spatial separation with increasing $y$ 
 can not be easily understood in terms of  {\em reversible coalescence} 
$A + A \rightleftharpoons A$~\cite{Avraham1990,Lin1992} (denoting each fascicle as a particle $A$).
The main reasons are: 
(i) the reverse reaction $A \to A+A$  allows for 
splitting of a single axon into two,  a mechanism 
not allowed in our model;
(ii) in contrast to reversible coalescence,  the probability of splitting of a fascicle in our model  decays rapidly with 
increasing fascicle size.
Thus the $y$-independent  distribution of inter-fascicle separation  $c_s \exp(-c_s \Delta x)$,
expected from IPDF of reversible coalescence~\cite{Avraham1990},
is never reached.
At large $y$, we find a distribution of inter-fascicle separation that conforms more to  ${\cal P}\Delta x\,\exp(-{\cal Q} \, \Delta x^2)$ (data not shown),
consistent with irreversible coalescence (see previous subsection).


We note again that time-dependent quantities in our model have no analog in the mapping to the 1d models we discussed above. The emergence of density-dependent long time scales in reversible coalescence~\cite{Avraham1990, Lin1992, Abad2002} therefore has no relation to the  long time scales in our model, which are due to a very slow reorganization of 
fascicle basins (Sec.~\ref{eff2}) --- a consequence of axon turnover.

}

\begin{figure*}[t]
\psfrag{S}{$S$}
 \psfrag{t/T}{$t/T$}
 \psfrag{(a)}{$(a)$}
 \psfrag{y=10       }{{$y=10~~~~~~$}}
 \psfrag{y=10^2       }{{$y=10^2~~~~$}}
 \psfrag{y=10^3       }{{$y=10^3~~~~$}}
 \psfrag{y=10^4       }{{$y=10^4~~~~$}}
  \psfrag{S(t)       }{{$S(t)$}}
\includegraphics[width=8.2cm]{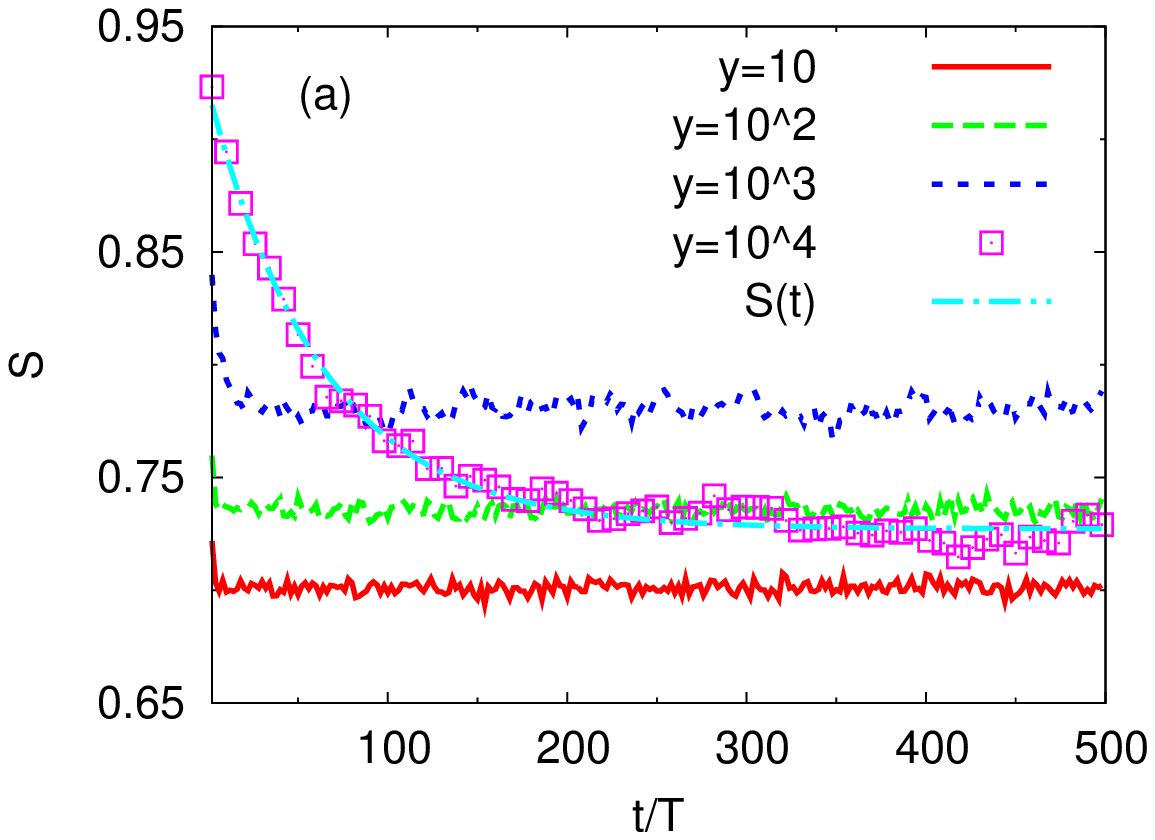}
\psfrag{S_inf}{$S_\infty$}
 \psfrag{y}{$y$}
 \psfrag{(b)}{$(b)$}
 \psfrag{ro=1/8       }{{$\r=1/8~~~~$}}
 \psfrag{ro=1/2       }{{$\r=1/2~~~~$}}
\includegraphics[width=8.2cm]{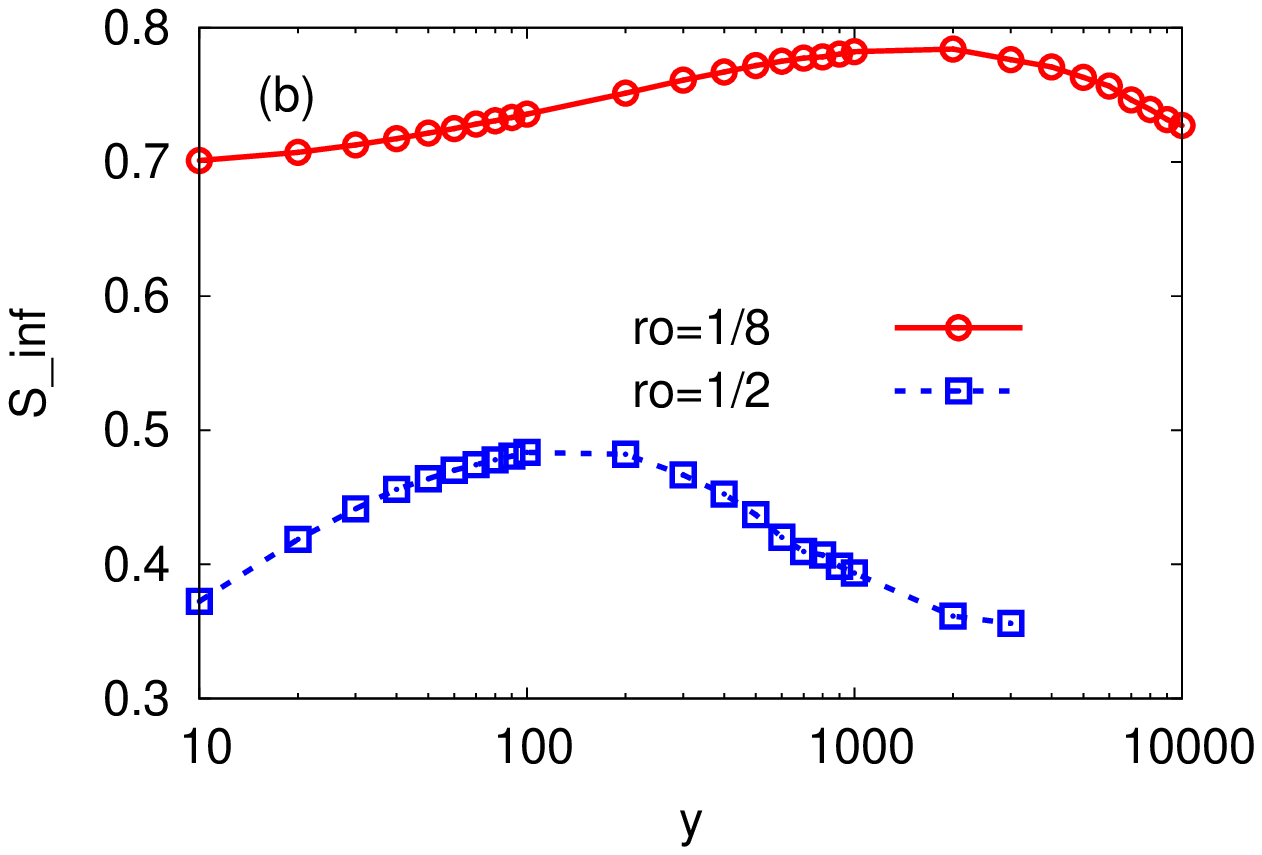}
\psfrag{n_inf}{$n_s$} 
 \psfrag{y}{$y$}
 \psfrag{(c)}{$(c)$}
 \psfrag{ro=1/8    }{{$\r=1/8~~~~$}}
 \psfrag{ro=1/2    }{{$\r=1/2~~~~$}}
 \psfrag{x^1/2    }{{$y^{1/2}~~~~$}}
\includegraphics[width=8.2cm]{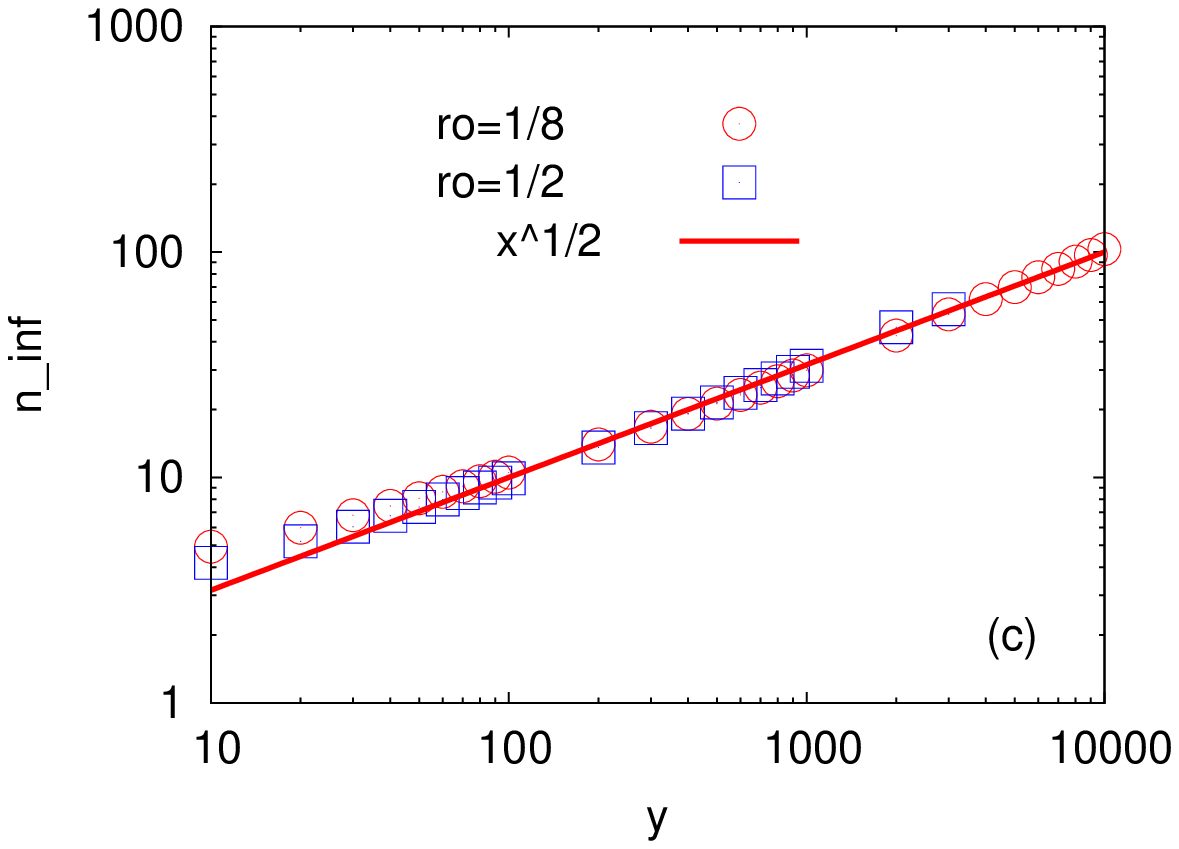}
\psfrag{tau}{$\t_{ap}/T$}
 \psfrag{y}{$y$}
 \psfrag{(d)}{$(d)$}
 \psfrag{t_s,ro=1/8    }{{$\r=1/8,\t_s~~~~$}}
 \psfrag{t_n,ro=1/8    }{{$\r=1/8,\t_n~~~~$}}
 \psfrag{t_s,ro=1/2    }{{$\r=1/2,\t_s~~~~$}}
 \psfrag{t_n,ro=1/2    }{{$\r=1/2,\t_n~~~~$}}
\includegraphics[width=8.2cm]{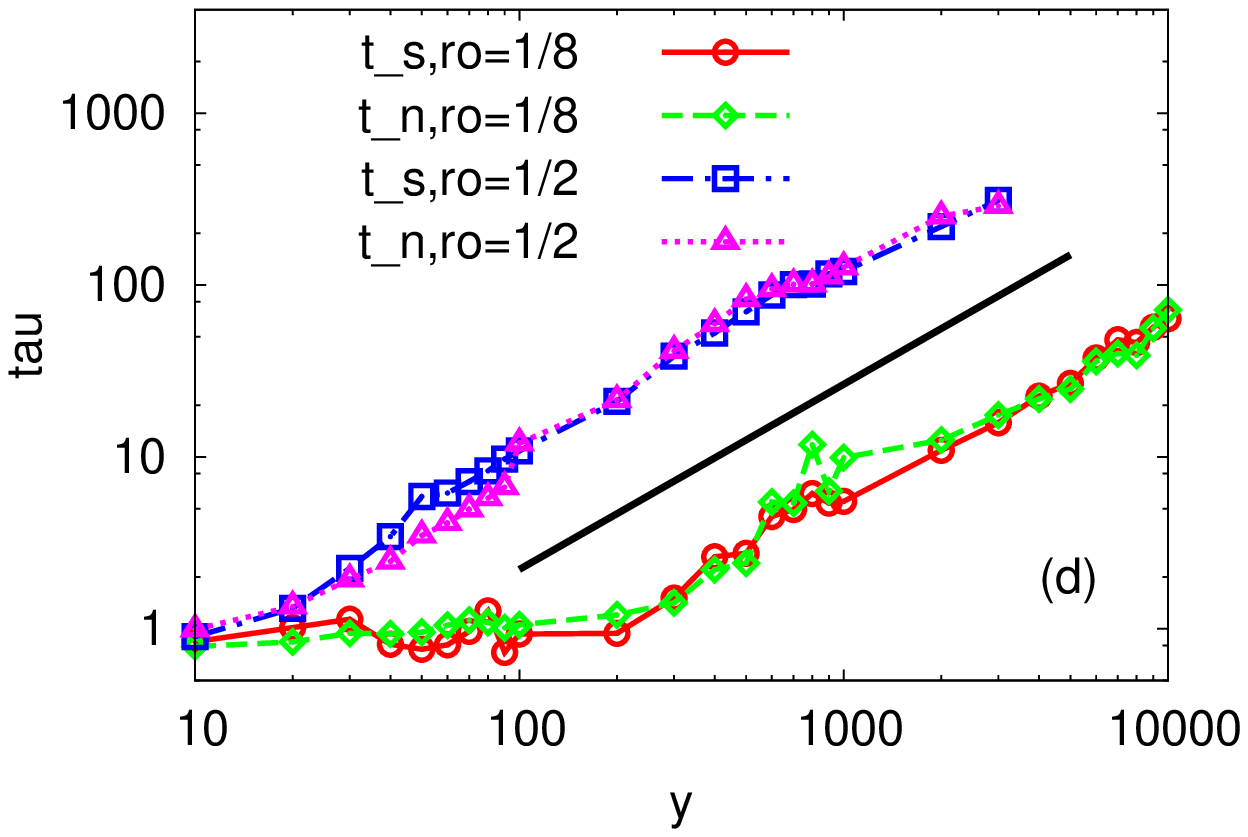}
\caption{(Color online) 
This plot shows the approach to steady state for mean fascicle size and mean purity
of fascicles in a system containing $N^r_0=50$ $r$-axons and $N^b_0=50$ $b$-axons
that interact through homotypic interaction strength $E_h=-4$ and heterotypic interaction
strength $E_o = -0.1$. Two mean axon densities $\r=1/8$ (system size $L=800$) and
$\r=1/2$ ($L=200$) are used.
($a$)~Approach-to-steady-state data for mean purity $S$ as a function of time $t/T$
in a system with density $\r=1/8$.
All the data were collected over $500T$ and averaged over $10^3$ realizations .
The data at each $y$-level fits to the form $S(t)=S_\infty+r\exp(-t/\t_s)$ where
$S_\infty$ is the asymptotic mean purity and $\t_s$ is the time scale of approach
to steady state. The fitting is shown for the data set at $y=10^4$,
where $S_\infty=0.73$, $r=0.19$ and $\t_s=63.64$ with all the fitting
errors being less than $2\%$.
($b$)~Asymptotic mean purity  $S_\infty$ as a function of 
$y$ at $\r=1/8$ and $\r=1/2$. 
{
Fitting errors in $S_\infty$ are within $3\%$.}
($c$)~Asymptotic subtracted mean fascicle size $n_s=(n_\infty-c)\exp(\be y)/2\r_{\rm{eff}}$ 
as a function of $y$. This follows $y^b$ with $b=1/2$. $\r_{\rm {eff}}$ is treated as a 
fitting parameter. For $\r=1/2$, $c=1.88 \pm 0.2$ and $\r_{\rm{eff}}=0.392 \pm 0.005$.
For $\r=1/8$, $c=1.13 \pm 0.05$ and $\r_{\rm{eff}}=0.059 \pm 0.001$.
($d$)~Approach-to-steady-state time scales, $\t_s$ for mean purity $S$
 and $\t_n$ for mean fascicle size $\bar n$,  as a function of $y$ for systems with 
 $\r=1/8$ and $\r=1/2$.
{
Fitting errors in $\t_s$ and $\t_n$ are within $3\%$.} 
 The time scales  
show an approximate power law growth $y^{2b}$ with $b=1/2$ 
denoted by the solid black line. Data are shown up to a $y$-level where the
time scales extracted from the fitting procedures remain less than half the total
run time $t=500 T$.}
\label{2sp-ap}
\end{figure*}
\begin{figure*}[t]
\psfrag{n_r}{$n^r$}
\psfrag{n_b}{$n^b$}
\psfrag{y=10}{$y=10$}
\psfrag{P(n_r,n_b)}{$P(n^r,n^b)$}
\includegraphics[width=8cm]{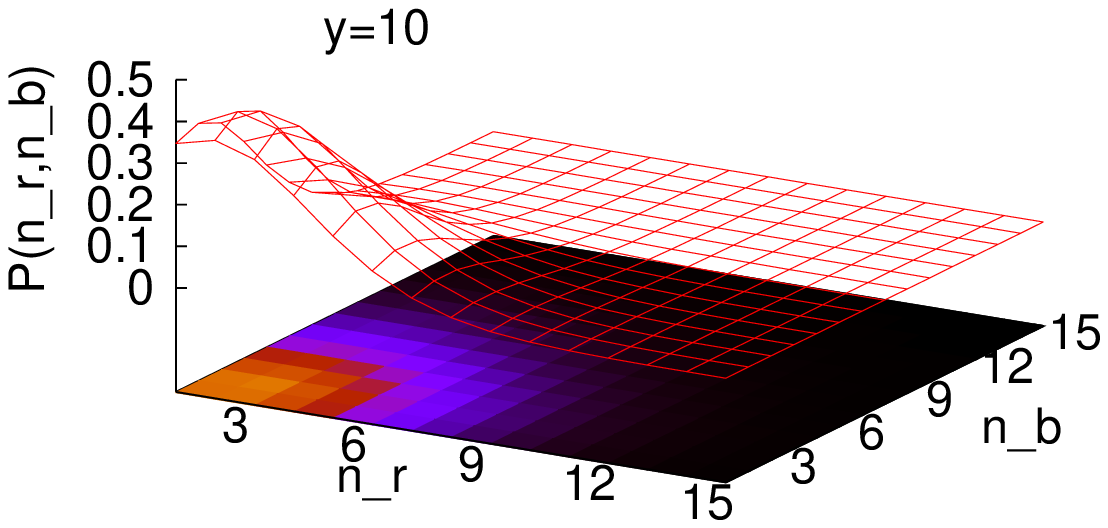}
\psfrag{n_r}{$n^r$}
\psfrag{n_b}{$n^b$}
\psfrag{y=100}{$y=10^2$}
\psfrag{P(n_r,n_b)}{$P(n^r,n^b)$}
\includegraphics[width=8cm]{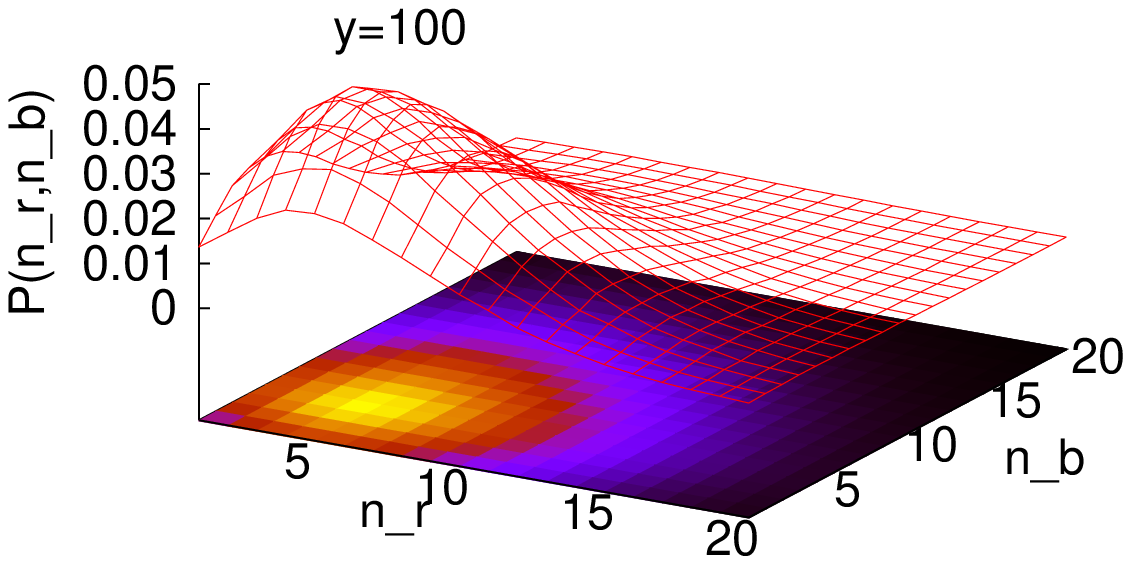}
\psfrag{n_r}{$n^r$}
\psfrag{n_b}{$n^b$}
\psfrag{y=10^3}{$y=10^3$}
\psfrag{P(n_r,n_b)}{$P(n^r,n^b)$}
\includegraphics[width=8cm]{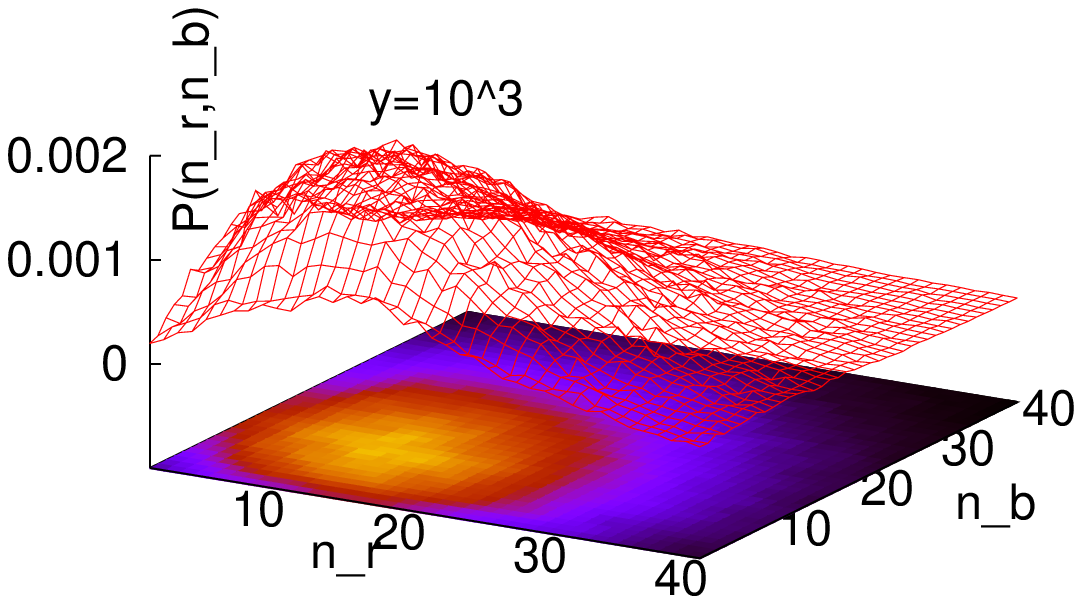}
\psfrag{n_r}{$n^r$}
\psfrag{n_b}{$n^b$}
\psfrag{y=10^4}{$y=10^4$}
\psfrag{P(n_r,n_b)}{$P(n^r,n^b)$}
\includegraphics[width=8cm]{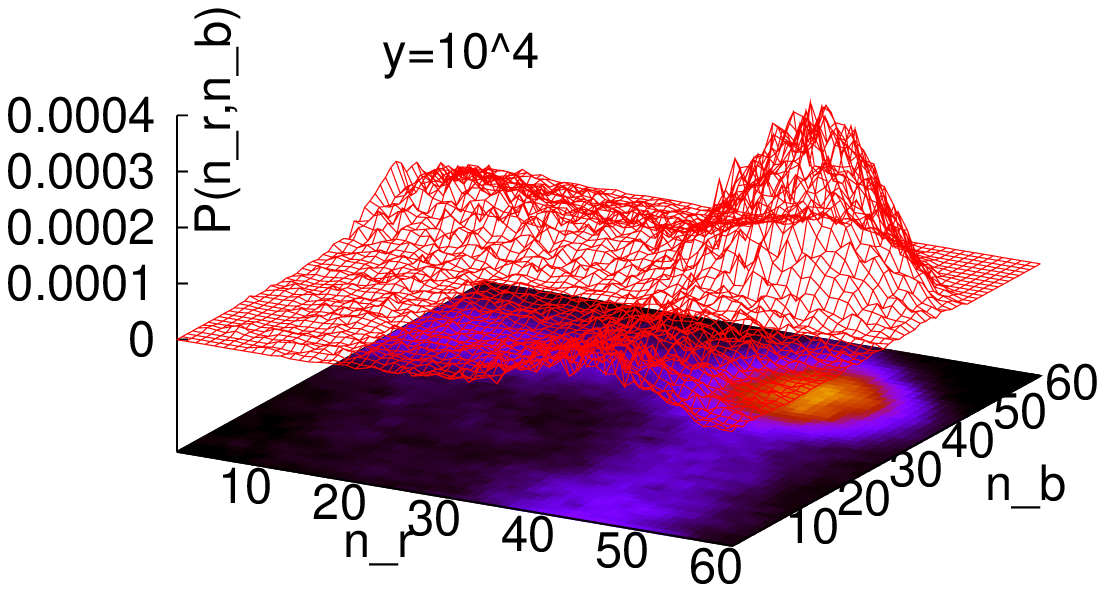}
\caption{(Color online) 
The distributions of fascicle composition $P(n^r,n^b)$ calculated by averaging over
$10^4$ realizations  and  time interval $10 T\leq t\leq 15 T$ in
a system with homotypic interaction $E_h=-4$,  heterotypic interaction  $E_o=-0.1$,  
number of axons $N^r_0=N^b_0=50$ and  system size $L=200$. The plots show
the distributions calculated at $y=10,\,10^2,\, 10^3,\, 10^4$.
}
\label{2spsteady}
\end{figure*}

\section{Mixed population of multiple axon types}
\label{2sp}
The axons of olfactory sensory neurons expressing 
distinct odorant receptors~\cite{Mombaerts} are believed to have 
short-range interactions 
with interaction strengths that are correlated with the
type of receptors the neurons express~\cite{homotypic1, homotypic2, homotypic3, Chehrehasa2006}.
In the framework of our model, this corresponds to the introduction 
of multiple types of random walkers, with type-dependent 
probabilities for attachment to / detachment from fascicles.
Neuronal systems containing multiple types of axons are known to achieve
pure and stable connections.
In the olfactory system, it may be expected~\footnote{Paul Feinstein, private communication.} 
that when the growth cone is located in a relatively pure environment
(i.e., when it is in contact with axons mainly of its own type), it leads to
reduced axonal turnover. While we do not include such effects in our model,
in this section we evaluate the mean purity of axon environment $S$ to
characterize the sorting dynamics.

\subsection{Mean purity}
Let us consider a system containing two types of axons named $r$ and $b$. 
Assume that a fascicle contains $n^r$ of $r$-axons and
$n^b$ of $b$-axons. If  the $i$-th axon in the fascicle is
of type $r$, the purity of environment that this axon encounters
within the fascicle is $s_i = (n^r-n^b)/(n^r+n^b)$, while if the axon is of
type $b$, the purity of environment is $s_i = (n^b-n^r)/(n^r+n^b)$. 
Then the mean purity of environment obtained by averaging over all $N(y)$ axons
is $S=(\sum_{i=1}^{N(y)} s_i)/N(y)$. The partial sum within a fascicle gives
$n^r(n^r-n^b)/(n^r+n^b)+n^b(n^b-n^r)/(n^r+n^b)=(n^r-n^b)^2/(n^r+n^b)$.
Thus, the degree of sorting  can be quantified as 
the mean purity of axon environment 
\bea
S=\f{1}{N(y)}\sum_{\mbox{fascicles}} \f{(n^r-n^b)^2}{(n^r+n^b)}.
\label{S}
\eea
Notice that $0\leq S\leq 1$;
$S=1$ corresponds to  completely pure fascicles containing only one type of
axons, whereas $S=0$ describes fascicles containing an equal mixture of the
two axon types.

\subsection{Approach to steady state}
\label{mean-2sp}
We consider a system of size $L=800$  having
$N^r_0=50$ axons of type $r$ and $N^b_0=50$
axons of type $b$ at the $y=0$ level ($\r=(N^r_0+N^b_0)/L =1/8$). 
We use the homotypic interaction energy
(interaction between $r$-$r$ or $b$-$b$)  $E_h=-4$ and the heterotypic
interaction energy (between the two different types $r$-$b$)  $E_o=-0.1$. We monitor the
time evolution of the mean purity of environment  $S$ and the mean fascicle
size $\bar n$. 
A typical configuration is shown in Fig.~\ref{conf}$(b)$.
The mean fascicle size reaches the steady state value at each
$y$-level with two characteristic time scales, 
similarly to the case of a system containing only one type of axons  (Fig.~\ref{ap}). 
The shorter time scale is intrinsic, the mean lifetime of a single axon $T$, 
and the other one is the emergent {
approach-to-steady-state} time scale $\t_n$ 
{
(equivalent to $\t_{ap}$ defined in the caption of Fig.~\ref{ap})}. 
As in the case of a system containing only one type of axons,
this time scale $\t_n$ can be orders of magnitude larger than $T$ (Fig.~\ref{2sp-ap}$(d)$).

The intrinsic time scale $T$ does not appear in the dynamics of $S$, however.
$S$ approaches its steady state value $S_\infty$ with a single time 
scale $\t_s$. 
The measured mean purity $S$ fits to the form
$S=S_\infty+r\exp(-t/\t_s)$ (Fig.~\ref{2sp-ap}($a$)). The
steady-state mean purity $S_\infty$ 
initially grows with $y$, reaching a maximum 
beyond which $S_\infty$ decreases (Fig.~\ref{2sp-ap}($b$)). 
This non-monotonic behavior is seen both for low ($\r=1/8$) and high
($\r=1/2$) density systems (Fig.~\ref{2sp-ap}($b$)). 
Thus for the combination of strong homotypic and weak heterotypic
attractive interaction, the fascicles achieve the highest purity
at a particular distance from their starting point, beyond which the
typical fascicle keeps on losing purity. From energetic considerations, one may understand
this behavior as follows. The high strength of the homotypic attraction compared to
the heterotypic one leads to sorting and thus the initial growth in purity
at lower $y$-levels.
Once one obtains highly pure and sufficiently large fascicles, however, the heterotypic interaction 
will  merge the $r$-dominated and $b$-dominated fascicles to form larger 
mixed fascicles, and thereby lead to a loss in mean purity.
The effect of the weak heterotypic interaction becomes significant only at higher
$y$-levels, where large fascicles are formed (recall that in our model, the interactions are additive).

Similar to the case of a system containing only a single axon type in presence of detachment, 
we find that the steady state mean fascicle size grows as
$n_\infty = c + 2\r_{\rm {eff}} y^{1/2} \exp(-\be y)$. 
Fig.~\ref{2sp-ap}($c$) shows that
the subtracted mean fascicle size $n_s = (n_\infty - c)\exp(\be y)/2\r_{\rm{eff}}$
follows the power law $y^b$ with $b = 1/2$. The effective density $\r_{\rm{eff}}$
and offset fascicle size $c$ are treated as fitting parameters. 

The time scales for approach to steady state $\t_n$ and $\t_s$ grow with $y$
following an approximate power law  $y^{2b}$ with $b= 1/2$ [Fig.~\ref{2sp-ap}($d$)]. 
This behavior is seen to be independent of density $\r$,
in contrast to the single-type case, where we found reliable $y^{2b}$ growth of the emergent time-scales
only at lower densities. Recall that, reduced inter-axon interaction strengths   
lead to lower effective densities $\r_{\rm{eff}}$ (Fig.~\ref{timePd}$(a)$). 
In the present case of mixed population of axons
of two types, the heterotypic interaction is very weak 
and may have lead to the effectively low-density ($y^{2b}$) power-law 
growth.
We note that $\t_n$ and $\t_s$ turn out  to be approximately equal 
to each other at all $y$-levels.

\subsection{Distribution of fascicle composition} 
In this section we briefly discuss the distribution of fascicle composition $P(n^r,n^b)$,
measured as the number of fascicles with $n^r$
$r$-axons and $n^b$ $b$-axons at a specified $y$-level and time $t$.
In Fig.~\ref{2spsteady} we show this distribution obtained by collecting data within 
the time interval 
$10 T\leq t\leq 15 T$ at various $y$-levels 
for a system of $N^r_0=N^b_0=50$ and $L=200$ ($\r=1/2$). From the 
approach-to-steady-state data (Fig.~\ref{2sp-ap}) it is clear that at $t=10T$, steady
state is reached only up to $y\approx 100$. Thus the distributions
obtained at $y=10^3,\, 10^4$ in Fig.~\ref{2spsteady} are far from 
steady state~\footnote{To obtain sufficiently good statistics for $P(n^r,n^b)$,
it was necessary to average over a large number of configurations. Consequently
we were restricted to a lower range of $t$ as compared to the data for mean fascicle
size $n$ and purity $S$ shown in Fig.~\ref{2sp-ap}.}. 
All the plots in Fig.~\ref{2spsteady} show a pronounced maximum for evenly
mixed fascicles, meaning that most of the fascicles we obtain are mixed by type.
However, a careful look at the plots reveals off-peak features of the distribution 
with reasonable weight
that reflect the presence of fascicles with highly asymmetrical composition 
(e.g.,  along the $n^b=1$ line 
for the plot at  $y=100$ in Fig.~\ref{2spsteady}, the maximum of the distribution 
is at $n^r=5$).

Finally, we comment on repulsive vs. attractive heterotypic interactions.
The attractive heterotypic interaction generates larger fascicles asymptotically,
but induces a reduction of purity at large  $y$. A repulsive heterotypic
interaction (combined with attractive homotypic interaction)
would  generate enhanced sorting, however it would be at the cost of a 
decreased mean fascicle size at all $y$ levels.

\section{summary}
\label{sum}
In this paper, we provided a simple model of the dynamics of axon fasciculation and sorting. To allow us to concentrate on the collective effects that arise from axon-axon interactions in a large population of axons, we chose a particularly simple implementation of single axon growth. In our model, each growing axon is represented as a directed random walk (Sec.~\ref{setup}). The common preferred growth direction may arise, e.g., from the influence of a spatially distributed guidance cue emitted by a distant target. Other than this common influence, we do not include in our model the guidance of axon growth cones by graded guidance cues, and restrict our attention to axon-axon interactions. The interaction of a growth cone with other axons is modeled as a short-range attractive interaction between a random walker and the trails of other random walkers (Sec.~\ref{setup}). In addition, we incorporated neuronal turnover (characteristic of, e.g.,  the mammalian olfactory system) by assigning a finite lifetime to each growing axon (Sec.~\ref{turnover}).

The strength of the axon-axon interaction was parametrized as an effective energy in a Monte Carlo update. In the energy-minimizing dynamics (corresponding to zero effective temperature in the Monte Carlo scheme), once a growth cone attaches to an axon fascicle, it will never detach (become a free random walker) again. For such dynamics, we extended our previous numerical and analytical results of Ref.~\cite{our_epl} (Sec.~\ref{onesp}). In the general dynamics (corresponding to unit effective temperature), axons may detach from fascicles, with a rate that increases with decreasing axon-axon interaction strength. We systematically studied how such detachment events modify the basic dynamics with no detachments (Sec.~\ref{onespD}). Finally, we investigated a system with two types of axons, and analyzed the sorting dynamics arising from a strong homotypic  interaction (between axons of same type) combined with a weak heterotypic interaction (between axons of different types) (Sec.~\ref{2sp}).
Our main findings were as follows.

The tendency to fasciculate is reflected in the growth of the mean fascicle size $\bar n$
with the distance $y$ in the preferred growth direction of the axons.
Using a mean field argument we showed that for the energy-minimizing dynamics 
the mean fascicle size $\bar n$ should grow  as 
$\bar n \simeq 2 y^{1/2}\, \r \exp(-\be y)$.  This agrees with the numerical results of Sec.~\ref{onesp}.
In Sec.~\ref{onespD} and \ref{2sp}, we showed that this growth law persists even in 
presence of  detachment,  with 
the average axon density $\r$ replaced by a fitting parameter  $\r_{\rm{eff}}$.

A more detailed characterization of the steady state is a position-dependent
fascicle size distribution.  Within  the scaling regime 
$L \ll y \ll (L/2)^2$, this distribution obeys a scaling law 
$P_s(n,y) = \la n(y) \ra^{-r} \phi(n/\la n(y)\ra)$ with 
$r=2.1$ and the scaling function
$\phi(u)= {\cal N} u \exp(-\nu u -\l u^2)$ (Sec.\ref{onesp}). 
At higher $y$-levels  the distribution becomes bimodal, a new 
maximum arises which is characteristic of the  complete fasciculation.
Even in the presence of  detachment, the scaling behavior of fascicle size distribution 
remains valid over a wide range of interaction strengths  
(Sec.~\ref{onespD} and \ref{2sp}).

The dynamics of reorganization of fascicles at high $y$-levels 
was found to be extremely slow. The emergent
time scales, e.g., the approach-to-steady-state  time $\t_{ap}$ or the auto-correlation 
time at steady state $\t_c$   can be orders of magnitude
larger than the mean lifetime of an axon $T$. 
In Sec.~\ref{eff2}, using an analytical model of  effective dynamics involving
two neighboring fascicles, we showed that the slowest mode of this dynamics
corresponds to the  exchange of basin space between the two fascicles, and grows 
with distance as $\t \sim y$. 
This behavior of time scales  survives even in the presence of detachment (shown in Sec.~\ref{meanD}).

{ 
While our model is two-dimensional, some limited analogies can be made to
one dimensional (1d) models of aggregation, coalescence, and chipping (Sec. VII).
We introduced a mapping in which the progressive fasciculation with
increasing $y$ (at a fixed time) in our model is mapped on to the time
evolution within a 1d system of interacting particles. Using
results from the literature on 1d models, we then obtained predictions
for stationary quantities in our model. Thus interpreting each axon as a
particle $A$ in the irreversible aggregation model $mA+nA \to (m+n)A$~\cite{benAvraham,redner}, we
obtained the prediction $u \exp(-\l u^2)$ for the distribution of fascicle sizes, which
is similar to the true steady-state distribution $u \exp(-\nu u-\l u^2)$ in our model with
energy-minimizing dynamics. Likewise, interpreting each fascicle of axons
as a particle A in the irreversible coalescence model $A+A \to A$~\cite{Avraham1990}, we obtained
the prediction $(\Delta x/y)\exp(-\mu \Delta x^2/y)$ for the distribution of separations between fascicles,
which agrees approximately with numerical results from our model (Sec.~\ref{1dcomp}). 
A limited analogy can also be made between the 1d
chipping (Ref.~\cite{mustansir,Rajesh2001}) or reversible coalescence (Ref.~\cite{Avraham1990,Lin1992}) models and our
model in the presence of detachment. Since in our model, the rate of
detachment decreases with the fascicle size, the reversible interaction
models are relevant only at low axon-axon interaction strengths. 
In this range of parameters, we were able to relate the
observed changes of distribution of fascicle sizes in our model to the
phase transition that occurs in the chipping model of Ref.~\cite{mustansir} (see Sec.~\ref{1dcomp}). 
We stress again, however, that the mapping to these 1d
models can say nothing about the time-dependent quantities in our model
(time in our model has no analog in the 1d models). Therefore, for
example, the slow time scales discussed in the reversible coalescence
model of Refs.~\cite{Avraham1990,Lin1992,Abad2002} are unrelated to the slow time scales present in our
model.
}

In Sec.~\ref{2sp} we analyzed a system with two types of axons
and type-specific interactions. 
In this system, axons sort into fascicles according to axon type.
We quantified the degree of sorting by introducing the mean purity $S$
of axon environment within fascicles. For the case of strongly attractive homotypic 
interaction and weakly attractive heterotypic interaction, 
we showed that  the degree of sorting $S$ varies with distance $y$ in a non-monotonic manner
and has a single maximum.

\section{Outlook}
\label{conc}
In this paper we analyzed a model aiming to describe the
formation of axon fascicles and the sorting of  fascicles by neuronal types  
in the mammalian olfactory system.
Our goal was to systematically investigate   the 
general non-equilibrium statistical mechanics aspects of the model,
leaving the task of building quantitative connections with physiology  for the future.

To conclude, we discuss possible generalizations of the basic model defined in this article,
 and the applicability to biological data on axon fasciculation. 
First, we note that in our discussion of the dynamical properties of the system,
it was essential that the random walkers moved in two (rather than three) spatial dimensions, 
and therefore cannot cross each other without interacting. In contrast, in a 
three dimensional system, the concept of fascicle basins would lose its validity. 
The resulting fascicle dynamics in three dimensions is expected to be significantly different
from the two-dimensional dynamics, which we showed to be governed by the 
competition of fascicles for basin space. 
It will be necessary to examine to what extent the assumption of two dimensionality 
is satisfied  in the olfactory system. 
However, this assumption  is effectively satisfied in studies of growth 
in neuronal cell culture~\cite{Honig98, hamlin,Voyiadjis2010}, in which the axons move 
on a plane surface, and interact when crossing each other. In Ref.~\cite{hamlin}, 
a fluorescence-based method is proposed for extracting the distribution of 
fascicle sizes; such experiments would permit a direct test of our model. 
{
Note that in some of these studies, our assumption of very strong adhesion of 
axons to the substrate is not satisfied. Events not included in our model, such as 
the gradual straightening of axon shafts or the local zippering / unzippering of 
fasciculated axons (as observed in Ref.~\cite{Voyiadjis2010}) may therefore occur.
}
Note also that natural boundary conditions
in such cell cultures are  either free boundary~\cite{hamlin} or 
confining channels~\cite{Honig98,Voyiadjis2010},
in contrast to the periodic boundary condition used in our simulations.

Recall that the mouse olfactory system contains about $1000$ axon types.
In this paper, we considered only up to $2$ types of axons. 
It would be interesting to examine if qualitatively new features emerge in 
systems with many axon types and a range of heterotypic interactions. 
This would require, however, significantly longer simulations.

Recently, Ref.~\cite{imai09} examined the role of axon-axon interactions in
the fasciculation and sorting of axons belonging to mouse olfactory sensory neurons.
This  showed that inter-axon repulsive interactions 
arising from Neuropilin-Semaphorin signaling play an important role in the axon sorting.
In the pre-target region (before reaching the olfactory bulb),
the amount of sorting grows with distance from olfactory epithelium~\cite{imai09}.
The pre-target axon sorting is shown to affect the topographic map formation by the neurons 
in the olfactory bulb~\cite{imai09,Bozza2009,Miller2010}. 
Using a mutant mouse,
Ref.~\cite{imai09} further showed that heterotypic axons sort even in absence
of the olfactory bulb, i.e., in complete absence of axon-target interactions.
This kind of experiments forms a suitable ground
for the application of our model. 
Notice that in our model, we have shown that a weaker heterotypic
attraction combined with a stronger homotypic attraction already leads to sorting. 
An effective repulsion between the two types would enhance the amount of sorting, however, 
it would be at the expense of the size of the fascicles formed.

Our immediate future goal is to extract the parameter values of our model from
controlled {\em in vitro} experiments and then use the model (and its possible extensions)
to analyze {\em in vivo} data on olfactory pattern formation
in mice.

\acknowledgments
We gratefully acknowledge extensive discussions with Paul Feinstein 
on olfactory development and on the formulation of our model. 
We thank James E. Schwob for communication on Ref.~\cite{hamlin}.
DC thanks FOM Institute AMOLF for support.
PB thanks the Pacific Institute for the Mathematical Sciences for partial support. 
MZ acknowledges support from the AV0Z50110509 fund and Center 
for Neuroscience LC554 fund (Czech Republic).

\bibliographystyle{apsrev4-1}

\begin{thebibliography}{52}
\bibitem [{why()}]{whyfascicle}%
  \BibitemOpen
  \href@noop {} {\bibinfo  {journal} {Throughout the text we use the term
  ``axon fasciculation'' rather than ``axon bundling'' as the term ``axon
  fasciculation'' is more standard in biological literature}}\BibitemShut
  {NoStop}%
\end{thebibliography}

%

\end{document}